\documentclass{aa}

\usepackage{graphicx}

\usepackage{txfonts}

\usepackage{enumitem}
\usepackage{adjustbox}
\usepackage{float}
\usepackage{hyperref}
\usepackage{xcolor, soul}
\usepackage{stfloats}
\usepackage{lipsum} 
\usepackage{capt-of}
\usepackage{booktabs}
\usepackage{tabularx}
\usepackage{makecell}
\usepackage{multirow}

\hypersetup{
    colorlinks=true,
    linkcolor=black,
    filecolor=magenta,      
    urlcolor=cyan,
    citecolor=blue
    }

\usepackage{makecell}

\usepackage[switch]{lineno} 

\begin{document}

   \title{Do little red dots really form a distinct class of astronomical objects?
}

\author{Jean-Baptiste Billand \inst{1} \thanks{jean-baptiste.billand@cea.fr}
    \and David Elbaz\inst{1}
    \and Maximilien Franco\inst{1}
    \and Fabrizio Gentile\inst{1,2}
    \and Emanuele Daddi\inst{1}
    \and Mauro Giavalisco \inst{3}
    \and Dale D. Kocevski \inst{4}
    \and Joseph S. W. Lewis \inst{1}
    \and Benjamin Magnelli\inst{1} 
    \and Valentina Sangalli \inst{1}
    \and Maxime Tarrasse\inst{5,6}
          }

\institute{
Université Paris-Saclay, Université Paris Cité, CEA, CNRS, AIM,  91191, Gif-sur-Yvette, France
\and INAF – Osservatorio di Astrofisica e Scienza dello Spazio, via Gobetti 93/3 - 40129, Bologna - Italy
\and University of Massachusetts Amherst, 710 North Pleasant Street, Amherst, MA 01003-9305, USA
\and Department of Physics and Astronomy, Colby College, Waterville, ME 04901, USA
\and School of Astronomy and Space Science, Nanjing University, Nanjing 210093, China 
\and Key Laboratory of Modern Astronomy and Astrophysics, Nanjing University, Ministry of Education, Nanjing 210093, China}

   \date{}

\abstract{
    \noindent Observations with the \textit{James Webb} Space Telescope have identified a previously unidentified class of enigmatic sources, known as little red dots (LRDs), which are relatively abundant in space. These sources have been interpreted as a distinct class of active galactic nuclei (AGNs) and host galaxies whose structural properties do indeed suggest the existence of a previously unidentified class of extragalactic objects. Such properties include the mass of the supermassive black hole and the emissivity from the associated AGN, as well as the stellar mass of the host galaxy or the possible presence of so-called quasi-stars or black hole stars (BH$^*$). However, there are two issues that have yet to be addressed before invoking a new paradigm: 1) whether there a clear discontinuity between the properties of the LRDs and field galaxies at the same cosmic epochs; and 2) whether LRDs form a single homogeneous population.
    In this work, we address these two issues by introducing a continuous metric to evaluate the "LRDness" of galaxies. We measured their compactness ($\delta_\text{compact}$), the sharpness of the V-shaped ($\delta_\text{v-shape}$) spectral energy distribution (SED), and the strength of the broad Balmer line emission. We applied this approach, which avoids imposing a binary “on--off” view of the population, to a sample of $\approx48,000$ ($\approx 5,000$) galaxies with photometric (spectroscopic) information, selected over an area of $\approx$750 arcmin$^2$. We find that the prominence of the V-shaped SED exhibits a strong correlation with morphology without any clear transition at the commonly used LRD selection threshold: the fraction of compact galaxies rises in line with the V-shape intensity. 
    Similarly, the strength of the H$\alpha$ broad line component increases with the V-shape sharpness and with compactness. We also note that the deficit of [N\,\textsc{ii}] itself is not an exclusive feature of LRDs, but a global property of compact, metal-poor galaxies.
    Finally, only a minority of LRDs (i.e., the 3 $\%$ most extreme ones) present a prominent Balmer break of potentially non-stellar origin (greater than 3). Overall, LRDs and non-LRDs show a similar Balmer decrement versus V-shape evolution, suggesting a common origin consistent with a dust attenuation origin. The inferred dust mass ($\approx 4-7 \times 10^4 M_\odot$) is low enough to explain ALMA non-detections, even if they are fully originating from stellar attenuation. The dust-driven origin is reinforced by the agreement between the observed Balmer ratios and attenuated Case B predictions.
    In conclusion, our results suggest that (apart from a minority, i.e., a few percent) LRDs do not generally represent a separate class of objects; rather, they are likely to make up the extreme tail of a continuous distribution of galaxies and broad H$\alpha$ emitters. Finally, we find that most LRDs appear to be consistent with a classical broad-line region and a dust component.
}


   \maketitle

\section{Introduction}

 The \textit{James Webb} Space Telescope (JWST) has opened up one of the most intriguing chapters in studies of the early Universe: the discovery of red, compact sources exhibiting broad Balmer lines \citep[e.g.,][]{labbe_population_2023, kocevski_hidden_2023, furtak_jwst_2023, kokorev_census_2024}, hereafter referred to as little red dots (LRDs; \citealt{matthee_little_2024}. These objects were initially interpreted as either dusty active galactic nuclei \citep[AGNs; e.g.,][]{kocevski_rise_2024, matthee_little_2024, greene_uncover_2024} or dusty, compact massive galaxies \citep{labbe_population_2023, Baggen_connecting, perez-gonzalez_what_2024}. However, subsequent observations have placed stringent constraints on their dust content. In particular, the flattening of the mid-infrared spectral energy distribution (SED) appears incompatible with the presence of a hot dust torus \citep{williams_galaxies_2024, perez-gonzalez_what_2024}. When combined with non-detections in the far-infrared, these limit the potential dust reservoir in LRDs \citep{akins_cosmos-web_2024, casey_dust_2024, xiao_nocII_2025, Setton_deficit_dust}. This evidence poses a challenge to both the highly obscured AGN and the dusty star-forming scenarios, where dust has previously been considered the primary driver of the observed red colors.

The discovery of unique sources, MOM-z-BH*1 and "the Cliff" \citep{Naidu_bhstar, degraaff_a_remarkable}, also challenged all previously proposed scenarios. These objects display Balmer breaks that are incompatible with a purely stellar component, and standard AGN templates hardly reproduce this feature \citep{Ma_uncover_404, degraaff_a_remarkable}. Similar sources were subsequently identified \citep{Taylor_2025_capers, barro_from}, and to explain such prominent Balmer breaks, an extremely dense gas reservoir along the line of sight has been invoked \citep{Ji_black_thunder, Inayoshi_2025}. Based on these observables, models featuring a black hole (BH) embedded in extremely dense gas, the so-called BH$^*$ model, have recently been proposed \citep[e.g.,][]{Naidu_bhstar, degraaff_a_remarkable, Rusakov_JWST_2025, Begelman_bh_star, Santarelli_quasi_star, Gentile_BHstar}. This scenario is rooted in the theoretical framework of quasi-stars originally developed by \citet{Begelman_2008}. In this framework, the dense layer surrounding the BH thermalizes and emits as a modified blackbody in the rest-frame optical and near-infrared. 

Furthermore, within the BH$^*$ model, broad lines arise not from a virialized broad-line region (BLR) as in the standard AGN paradigm \citep[e.g.,][]{Greene_BH_mass, Reine_BH_AGN}, but through physical processes such as electron or resonance scattering \citep[e.g.,][]{Naidu_bhstar, torralba_Z_LRD, Rusakov_JWST_2025, Chang_lines_bhstar, Gentile_BHstar}. Depending on the model, the inferred BH masses can be up to two orders of magnitude lower than those typically inferred or, for instance, as in \citet{Gentile_BHstar}, the BH masses are still massive compared to the host galaxy. However, several LRDs appear to be inconsistent with this picture. For instance, a dynamical BH mass was measured for one source \citep{Juodzbalis_bh_mass}, showing agreement with the mass estimated from the H$\alpha$ line. Additionally, a multi-line analysis of the "Rosetta Stone" LRD was shown to disfavor resonance and electron scattering as the primary mechanism for line broadening \citep{Juodzbalis_jades_agn,Brazzini_ruling_out}. While the BH$^*$ model predicts the X-ray weakness observed in most LRDs \citep{Begelman_bh_star, Paccucci_direct_collapse, Kido_bh_star}, several LRDs have now been detected in X-rays. These observations support the idea that not all LRDs can be classified purely as BH$^*$ objects as proposed, indicating that a more complex physical picture is required \citep{kocevski_rise_2024,Fu_LRD_xray, Hviding_XRD}.

Moreover, the definition of LRDs varies depending on the applied selection criteria, leading to classification ambiguities for sources near the boundaries. Consequently, an object’s classification as an LRD may depend solely on specific sample limits. For instance, various color-color selections have been developed \citep[e.g.,][]{kocevski_rise_2024, kokorev_census_2024, akins_cosmos-web_2024, barro_photometric_2024}, each introducing slightly different thresholds that result in significantly different sample compositions. Models built on solely conservative selections likely miss less extreme objects, creating a biased view of the population. This implies that broad-line Balmer emitters located near these selection boundaries can be interpreted as BH$^*$ objects or as typical AGNs, which are more massive by two orders of magnitude, depending on arbitrary selection cuts.

This raises a fundamental question, based purely on observables, regarding whether LRDs represent a distinct class of astronomical objects. To address this, we propose a new metric to characterize the features of LRDs, specifically their compactness and V-shaped SED. Unlike traditional color-color cuts, this approach avoids arbitrary binary thresholds by treating these characteristics as continuous parameters, $\delta_{\text{compact}}$ and $\delta_{\text{v-shape}}$.

We investigate in this work, using this continuous approach, whether LRDs exhibit discontinuities in their observed properties with galaxy fields, which would make appear a natural selection for them and signifying a separate population, or whether they demonstrate continuous behavior, suggesting instead that they form a continuum with the broader galaxy population. To address this question, we combined photometric and spectroscopic data over the four widest fields covered by the JWST: PRIMER-UDS, PRIMER-COSMOS, JADES (GS+GN), and CEERS, as described in Sect. \ref{sec:data}. In Sect. \ref{sec:lrdness} we introduce a continuous approach to quantify the main features of LRDs. Section \ref{sec:Halpha_evolution} focuses on the behavior of the H$\alpha$ emission line and we discuss the physical interpretation of our findings within the LRD framework in Sect. \ref{sec:discussion}. Magnitude used are in the AB system \citep{OkeJB_abmag} and the cosmological parameters from \citet{Planck_collaboration_cosmo_param}.

\section{Data}\label{sec:data}

\subsection{Photometric data}\label{sec:photometric_data}
In this work we utilized publicly available photometric catalogs from several JWST surveys. We specifically selected fields providing coverage in at least the following six NIRCam wide-band filters: F115W, F150W, F200W, F277W, F356W, and F444W. These data were supplemented by the \textit{Hubble} Space Telescope observations in the F606W and F814W bands from ACS/WFC \citep{Davis_HST_2007, Ford_1998_hubble}. This comprehensive multi-band coverage enabled an accurate determination of both rest-frame optical and UV slopes between $2<z<10$, as detailed in Sect. \ref{sec:lrdness}. The surveys used in this study are described below:

\begin{itemize}
    \item JWST Advanced Deep Survey \citep[JADES; ][]{Eisenstein_jades,Hainline_phot_z_jades, Rieke_phot_z_jades}: We used the official DR2 and DR3 photometric catalogs for the GOODS-N and GOODS-S fields, covering approximately 175~arcmin$^2$ and F444W$_{5\sigma \text{ point-source}}=29.62$. Photometric redshifts provided in these catalogs were computed using \texttt{EAZY} \citep{Eazy_z}, with $\sigma_\text{NMAD}$\footnote{The parameter $\sigma_{\text{NMAD}}$ represents the scatter in the photometric versus spectroscopic redshift relation, defined as $1.48 \times \text{median} \left( \frac{|\Delta z - \text{median}(\Delta z)|}{1 + z_{\text{spec}}} \right)$, where $\Delta z = z_{\text{phot}} - z_{\text{spec}}$.}$\approx 0.037$.

    \item Cosmic Evolution Release Science \citep[CEERS;][]{Finkelstein_CEERS, Finkelstein_CEERS2}: Located in the EGS field and covering 97~arcmin$^2$ and F444W$_{5\sigma \text{ point-source}}=28.97$. We adopted the official catalog from \citet{Cox_CEERS_catalog}, which includes both photometry and photometric redshifts determined with \texttt{LePHARE} \citep{Arnouts_phot_z, Ilbert_phot_z}, with $\sigma_\text{NMAD}\approx 0.035-0.073$.

\item Public Release IMaging for Extragalactic Research \citep[PRIMER;][]{Dunlop_primer}: This survey covers $\approx 500$~arcmin$^2$ across the COSMOS and UDS fields (PRIMER-COSMOS and PRIMER-UDS) with F444W$_{5\sigma \text{ point-source}}=28.2 - 28.5$. We used the photometric catalogs provided by \citet{sun_PRIMER} along the photometric redshift computed with EAZY, $\sigma_\text{NMAD}\approx 0.027$.
\end{itemize}

The catalogs already include spectroscopic redshifts when available, which we supplement with data from the DAWN JWST Archive (DJA; v4.4). More details are provided in Sect. \ref{spectroscopic_data}. To ensure robust photometric detections and reliable redshift estimates, we excluded sources with a signal-to-noise ratio below 10 in F444W  ($S/N < 10$). Furthermore, for sources lacking spectroscopic redshifts, we removed those with poorly constrained photometric redshifts with $\chi^2_{\text{phot}} > 15$.

\subsection{Spectroscopic data}\label{spectroscopic_data}

Throughout this work, we utilized the spectroscopic data available on DJA, the NIRSpec microshutter array (MSA) spectra table (version 4.4). The spectra were reduced using the \texttt{msaexp v0.9.8} pipeline  \citep{Brammer2022msaexp}, following the methodology described in \citet{Heintz_JWST_spectra} and \citet{deGraaff_RUBIES}, see \citet{DeGraaff_BHstar} for further details. For PRISM data, we exclusively used spectra with a \texttt{grade = 3}, ensuring high-quality data and forming a sample of approximately 17,000 spectra from numerous surveys\footnote{See \href{https://dawn-cph.github.io/dja/spectroscopy/nirspec/}{dawn-cph/dja/spectroscopy/nirspec} for a complete list.}, including CEERS \citep{Finkelstein_ceers_spectra}, JADES \citep{Eisenstein_jades,deugenio_spectra,curtis_lake_jades_spectra,Scholtz_jades_spectra}, NEXUS \citep{Shen_NEXUS}, UNCOVER \citep{Bezanson_UNCOVER,Price_uncover}, GTO-Wide \citep{Maseda_GTO_wide}, CAPERS (GO-6368 and PI Dickinson), and RUBIES \citep{deGraaff_RUBIES}.

Next, we merged the PRISM data with our photometric catalog (by cross-matching at 0.1 arcsec), including the spectroscopic redshifts where possible. For the analysis in Sect. \ref{sec:Halpha_evolution}, we also used the grating data provided by the DJA.

\section{Evaluating the ``LRDness'' in galaxies: $\delta_\text{compact}$ and $\delta_\text{v-shape}$}\label{sec:lrdness}

To determine whether LRDs represent a distinct population with unique observed properties compared to other galaxies or whether they lie along a continuum, exhibiting progressively more extreme properties,  we adopted an approach that is as agnostic as possible regarding their selection. Specifically, we eliminated any a priori assumptions about their properties and selection criteria. To achieve this, we constructed a continuous scale applicable to all galaxies, based on two key parameters: $\delta_\text{compact}$ and $\delta_\text{v-shape}$. The first parameter, $\delta_{\text{compact}}$, quantifies the galaxy's compactness. The second, $\delta_{\text{v-shape}}$, measures the prominence of the V-shaped SED, characterized by the combination of a red rest-frame optical continuum and a blue rest-frame UV continuum. These two parameters capture the two primary characteristics of LRDs: a V-shaped SED and a compact morphology, but on a continuous scale. The spectroscopic analysis of the H$\alpha$ line, the third key property of LRDs, is done a posteriori in Sect.~\ref{sec:Halpha_evolution}.

\begin{figure}[h!]
    \centering
    \includegraphics[width=9cm]{ 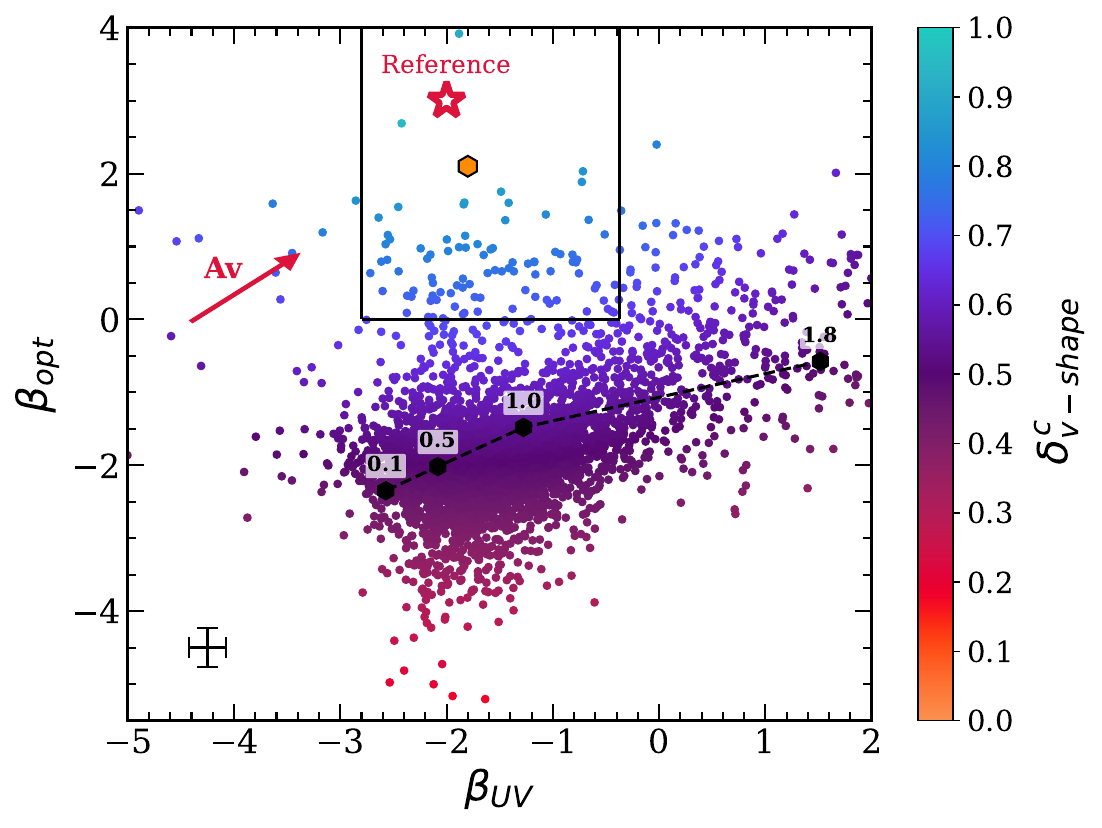}
    \caption{Distribution of sources in the $\beta_{\text{UV}}$--$\beta_{\text{opt}}$ plane, illustrating the $\delta_{\text{v-shape}}$ parameter. Higher values of $\delta_{\text{v-shape}}^c$ (approaching 1) indicate a more prominent V-shaped SED. The selection criteria from \citet{kocevski_rise_2024} are overlaid as solid black lines for comparison. The red star represents the reference point $(\beta_{\text{UV,ref}} = -2, \beta_{\text{opt,ref}} = 3)$ used to derive $\delta_{\text{v-shape}}$. A typical error bar for the $(\beta_{\text{UV}}, \beta_{\text{opt}})$ measurements is shown as a cross in the bottom-left corner. The red vector illustrates the effect of increasing dust attenuation on the stellar population, assuming a \citet{Calzetti_2000} law. The dotted line represents the evolutionary track of a dust-free galaxy, assuming a delayed star formation history with a fixed $\tau = 200$ Myr and a stellar age evolving from $0.1$ to $1.8$ Gyr. For reference, we include CEERS 746 (orange hexagon), a broad-line AGNs studied by \citet{kocevski_hidden_2023}, whose optical colors can be explained by either a dusty starburst galaxy or a obscured quasar.}
    \label{fig:delta_vshape_primer_uds} 
\end{figure}

\subsection{Intensity of the V-shape: $\delta_\text{v-shape}$}

Typically, LRDs are selected using thresholds in either color-color or color-magnitudes diagrams \citep{labbe_uncover_2023,greene_uncover_2024,perez-gonzalez_what_2024,kokorev_census_2024,barro_photometric_2024,akins_cosmos-web_2024,rinaldi_not_2024} or based on their rest-frame optical--UV slope, $\beta_\text{UV}$ and $\beta_\text{opt}$, which presents a blue UV continuum and a red optical continuum \citep{kocevski_rise_2024,leung_exploring_2024,Hviding_RUBIES,DeGraaff_BHstar}. However, as these thresholds are arbitrary, the delimitation and definition of LRDs remain non-unified and strongly depend on the authors and available data. Consequently, objects near the selection boundary are sensitive to the method applied. For instance, the color-color selection of \citet{barro_photometric_2024} tends to include more objects than that of \citet{kocevski_rise_2024}.

This raises a fundamental question regarding these "less extreme" candidates, which are missed by conservative selection criteria, namely, whether they can be interpreted as members of the LRD population, whether they exhibit the same properties, or whether they belong to another class. To address this fundamental question, we introduce the following parameter: $\delta_\text{v-shape}$.

To this end, we computed the rest-frame UV slope, $\beta_\text{UV}$, and the rest-frame optical slope, $\beta_\text{opt}$, as described in \citet{kocevski_rise_2024}, using the photometric data presented in Sect. \ref{sec:photometric_data}. These slopes are defined by the relation $f_\lambda \propto \lambda^\beta$, where the slopes on the blue side correspond to $\beta_\text{UV}$ and slopes on the red side to $\beta_\text{opt}$, measured on either side of the Balmer break at 3645\,\AA. At different redshift intervals, the set of bands switch accordingly to always probe the two regions of interest. Between $2<z<8$, both the UV slope and optical slopes are computed with 3 filters each, whereas at a higher redshift of $z>8$, the optical slope is computed only with two bands (see \citet{kocevski_rise_2024} for more details). We then define the parameter $\delta_\text{v-shape}$ as

\begin{eqnarray}
      \delta_\text{v-shape} = 1 -  \frac{\sqrt{(\beta_\text{UV} - \beta_\text{UV,ref})^2 + (\beta_\text{opt}-\beta_\text{opt,ref})^2}}{10} \ ,
\end{eqnarray}

\noindent with $(\beta_\text{UV,ref}, \beta_\text{opt,ref})$ representing a reference point in the $\beta_\text{UV}, \beta_\text{opt}$ plane. The parameter $\delta_\text{v-shape}$ is therefore defined as the Euclidean distance from an arbitrary point, normalized to a maximum distance (here, 10). This reference point is chosen to correspond to an extreme V-shape: $(\beta_\text{UV,ref} = -2, \beta_\text{opt,ref} = 3)$. When possible, using the NIRSpec MSA v4.4 table, we corrected the optical slope for the H$\alpha$ flux contribution (as detailed in Appendix \ref{sec:appendix0}). In the following, we make a distinction between our photometric sample and the spectroscopic one. For the latter, we have a spectroscopic redshift, a corrected optical slope, and a corrected $\delta_\text{v-shape}$ (hereafter denoted as $\delta_\text{v-shape}^c$). Typically, the correction applied to sources with $\delta_{\text{v-shape}} > 0.6$ is approximately $0.01$. The $\delta_{\text{v-shape}}$ intervals used in the figures throughout this work are at least four times larger than this value, ensuring that the observed trends remain independent of this correction. To ensure robust measurements of both the rest-frame optical and UV slopes across the different fields used in this work, as well as homogeneity between our spectroscopic and photometric samples, we applied a magnitude cut in the filter covering the H$\alpha$ line. Depending on the redshift range, this filter varies between F277W, F356W, and F444W. Specifically, we require $m_{\text{H}\alpha} < 27.5$ (see Appendix \ref{sec:appendix0} for more details).

An illustration of $\delta_\text{v-shape}^c$ is provided in Fig.~\ref{fig:delta_vshape_primer_uds} for our spectroscopic sample across all the fields. We also display a fiducial dust-free galaxy track. This track was generated using \texttt{CIGALE} \citep{cigale} to model the SED, assuming a delayed star formation history with a fixed e-folding time of $\tau = 200$~Myr and a stellar population age evolving from $0.1$ to $1.8$~Gyr. The impact of dust reddening is represented by a red vector, calculated assuming a \citet{Calzetti_2000} attenuation law. For comparison, we included CEERS~746, an LRD previously studied by \citet{kocevski_hidden_2023}. The optical SED of this source is degenerate, as can be well-described by either a reddened Type~1 quasi-stellar object or a dusty starburst galaxy.

\begin{figure}[h!]
    \centering
    \includegraphics[width=7cm]{ 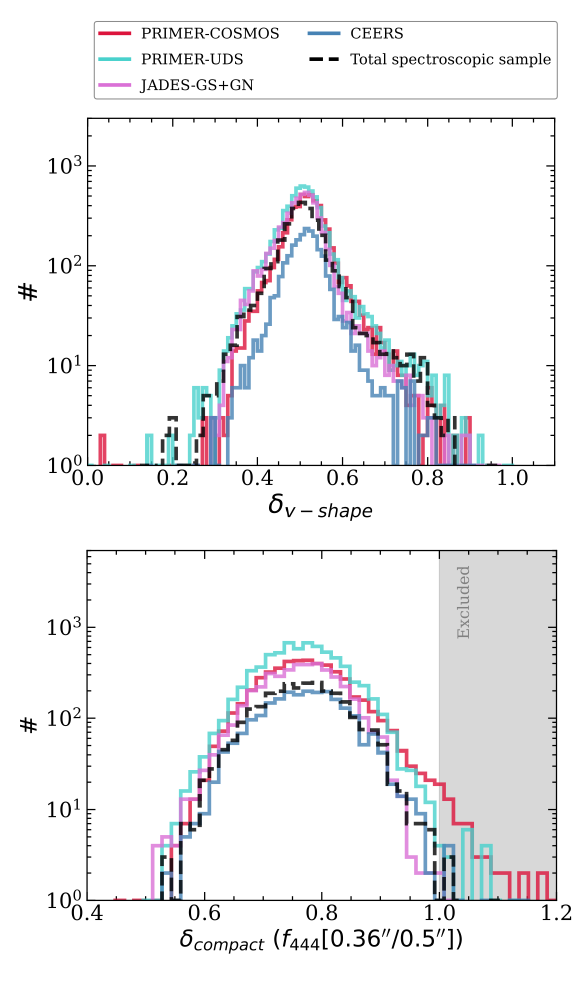}
    \caption{\textit{Top}: Distribution of the $\delta_\text{v-shape}$ parameter for every field used in this work. As $\delta_\text{v-shape}$ approaches 1, the more extreme the V-shaped SED. \textit{Bottom}: Distribution of  $\delta_\text{compact}$ in every fields use in this work, as $\delta_\text{compact}$ increases, galaxies are more compact, point source dominated. In shaded gray we show the excluded region to avoid artifacts. The dashed black line denote our main spectroscopic sample across all the fields. 
}

    \label{fig:delta_v_shape}
\end{figure}

This parameter offers the advantage of being a continuous scale that evaluates the intensity of the V-shape, without excluding less extreme objects. Based on this scale, a galaxy with a $\delta_\text{v-shape}$ value close to one exhibits an extreme V-shaped SED, whereas the majority of galaxies present a less intense V-shaped SED, with $\delta_\text{v-shape} \approx 0.5$, as shown by the top panel in Fig.~\ref{fig:delta_v_shape}. The distribution remains consistent across all fields, indicating that the measurement of this parameter is independent of the field used, thus enabling reliable comparisons between fields.

\subsection{Compactness: $\delta_\text{compact}$ }

To quantify the compactness of the galaxies in our sample, we utilized the public data from the DJA. By cross-matching our photometric catalog across all fields (at $0.1''$), we incorporate the aperture photometry provided by the DJA in the F444W filter. Following a similar methodology as in \citet{greene_uncover_2024, kokorev_census_2024, akins_cosmos-web_2024, barro_photometric_2024}, we employ an aperture flux ratio to define source compactness. Since this approach is model-independent, it avoids the convergence issues that can appear in 2D S\'ersic fits for sources near the resolution limit, particularly in the case of LRDs. Given that the specific apertures provided by the DJA differ from those used in the literature, we define our compactness parameter, $\delta_{\text{compact}}$, as

\begin{equation}
    \delta_{\text{compact}} = \frac{f_{444}(d=0.36'')}{f_{444}(d=0.5'')} \ .
\end{equation}

Similarly to the case of the $C_{444}$ parameter used by \citet{akins_cosmos-web_2024}, this ratio can be sensitive to artifacts or hot pixels at extreme values. To mitigate these effects, we clipped the $\delta_{\text{compact}}$ distribution at the same sigma-level from the median to match the threshold applied by \citet{akins_cosmos-web_2024}. In our sample, this corresponds to a cut at $\delta_{\text{compact}} = 1$; therefore, we excluded any sources with values exceeding this limit. Furthermore, to minimize the redshift dependence of $\delta_{\text{compact}}$ (which is based solely on the F444W band), we restricted our analysis to $3 < z < 9$. This ensured that we would be probing the rest-frame compactness within the $\approx 0.5$--$1.1~\mu$m range. This parameter utilizes fluxes within aperture diameters of $0.36''$ and $0.5''$. While this differs from the apertures commonly used in the literature (typically $0.2''$ and $0.4''$ or $0.5''$), we draw a comparison between $C_{444}$ and $\delta_{\text{compact}}$ in Appendix \ref{sec:appendix1} and show that both are strongly correlated. Therefore, $\delta_{\text{compact}}$ is also an efficient proxy to probe compact objects. By cross-matching our catalog with the morphological measurements provided by the DJA \citep[using a $0.1''$ radius;][]{Genin_DJA}, we derived a median half-light radius of $R_e = 0.076^{+0.10}_{-0.05}$~arcsec for sources with $\delta_{\text{compact}} > 0.8$. This median value lies below the half-width at half-maximum of the F444W point spread function ($\approx 0.08$~arcsec), confirming that the majority of the sources above this compactness value are point source-dominated. We note that a similar distribution is observed across the different fields in our spectroscopic sample for both the $\delta_\text{v-shape}$ and $\delta_\text{compact}$ parameters (see Fig. \ref{fig:delta_v_shape}), as for the photometric sample.

\begin{figure}[h!]
    \centering
    \includegraphics[width=9cm]{ 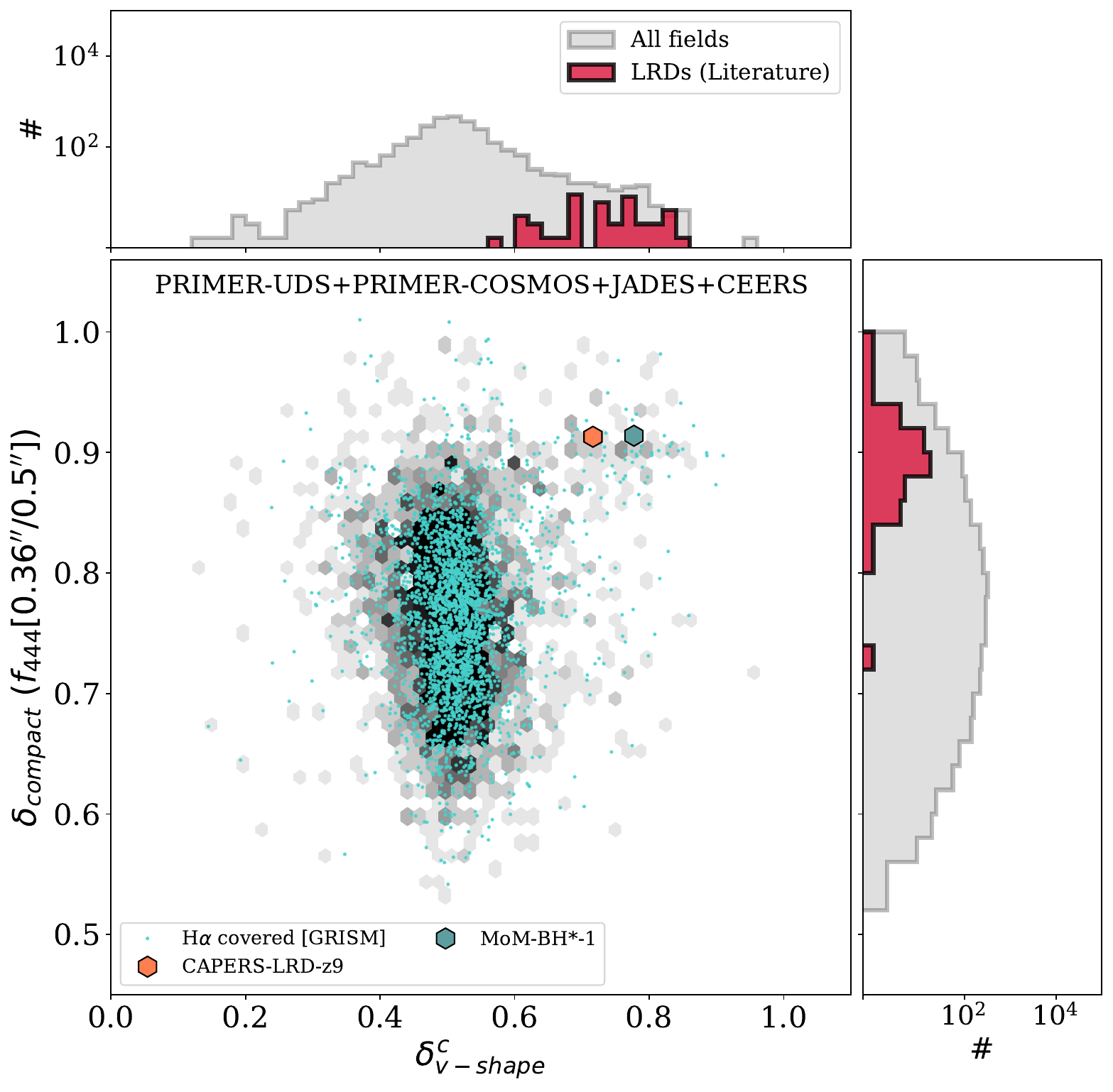}
    \caption{ $\delta_\text{compact}$ as a function of $\delta_\text{v-shape}^c$, for  all the galaxies in every fields studied: PRIMER-UDS, PRIMER-COSMOS, JADES-GS, JADES-GN, and CEERS. Cyan points correspond to galaxy for which H$\alpha$ covering from at least grating medium resolution is available. Cyan and orange hexagon represent Mom-BH*-1 and CAPERS-LRD-z9 from \citet{Naidu_bhstar} and \citet{Taylor_2025_capers}, respectively. In the corner plots, the red histograms represent the merged LRD sample from \citet{kocevski_rise_2024} and \citet{akins_cosmos-web_2024}, while the gray histograms show the distribution of the entire galaxy population.
}
    \label{fig:delta_vshape_delta_compact}
\end{figure}

\subsection{Summary of the sample selection}\label{sec:sample_selection}
We excluded sources with poorly constrained photometric redshifts, defined as those with $\chi^2_{\text{phot}} > 15$ when the spectroscopic redshifts are not available. Here, we summarize the remaining steps and selection criteria we employed to formulate our final sample: 

\begin{itemize}
    \item We required the S/N in F444W to be above 10.
    \item We required the filter covering the H$\alpha$ line to be brighter than $27.5$ AB mag (i.e., $m_{\text{H}\alpha} < 27.5$).
    \item Brown dwarfs were removed by applying an UV slope cut of $\beta_{\text{UV}} > -2.9$.
    \item Sources with $\delta_{\text{compact}} > 1$ were excluded.    \item The sample was restricted to  $3 < z < 9$.
\end{itemize}

Our final photometric sample consists of 47,928 sources, of which 7,072 have spectroscopic redshifts. Our spectroscopic sample comprises 5,185 sources for which we corrected the v-shape intensity for the H$\alpha$ contribution using PRISM data. An additional 2,960 sources have higher resolution (medium or high grating) spectra covering the H$\alpha$ region. Most of the following figures in this study are restricted to our spectroscopic sample.

\subsection{Correlation between $\delta_\text{v-shape}$ and $\delta_\text{compact}$ \label{subsec:corralation_dvshape_dcompact}}

The first aspect we investigate here is the potential correlation between the V-shaped SED in galaxies and their compactness, combining data from all fields. In Fig.~\ref{fig:delta_vshape_delta_compact} we first present $\delta_\text{compact}$ as a function of $\delta_\text{v-shape}$, along with the available spectroscopic data (i.e., medium and high resolution that covered H$\alpha$ used in Sect.~\ref{sec:Halpha_evolution}).

By cross-matching the LRDs sample of \citet{kocevski_rise_2024} and \citet{akins_cosmos-web_2024}, we determine typical values for LRDs to be $\delta_\text{v-shape} > 0.6$ with $\delta_\text{compact} > 0.8$. As expected, typical LRDs are located in the top-right region of the diagram. For reference, we also highlight two extreme LRDs\footnote{We note that {The Cliff} from \citet{degraaff_a_remarkable} is blended with a foreground source in F444W, making the measurement of $\delta_\text{compact}$ not relevant.} in Fig.~\ref{fig:delta_vshape_delta_compact}, \textit{CAPERS-LRD-z9} and \textit{MoM-BH*-1} \citep{Taylor_2025_capers,Naidu_bhstar}.

\begin{figure*}[t!]
    \centering
    \includegraphics[width=17cm]{ 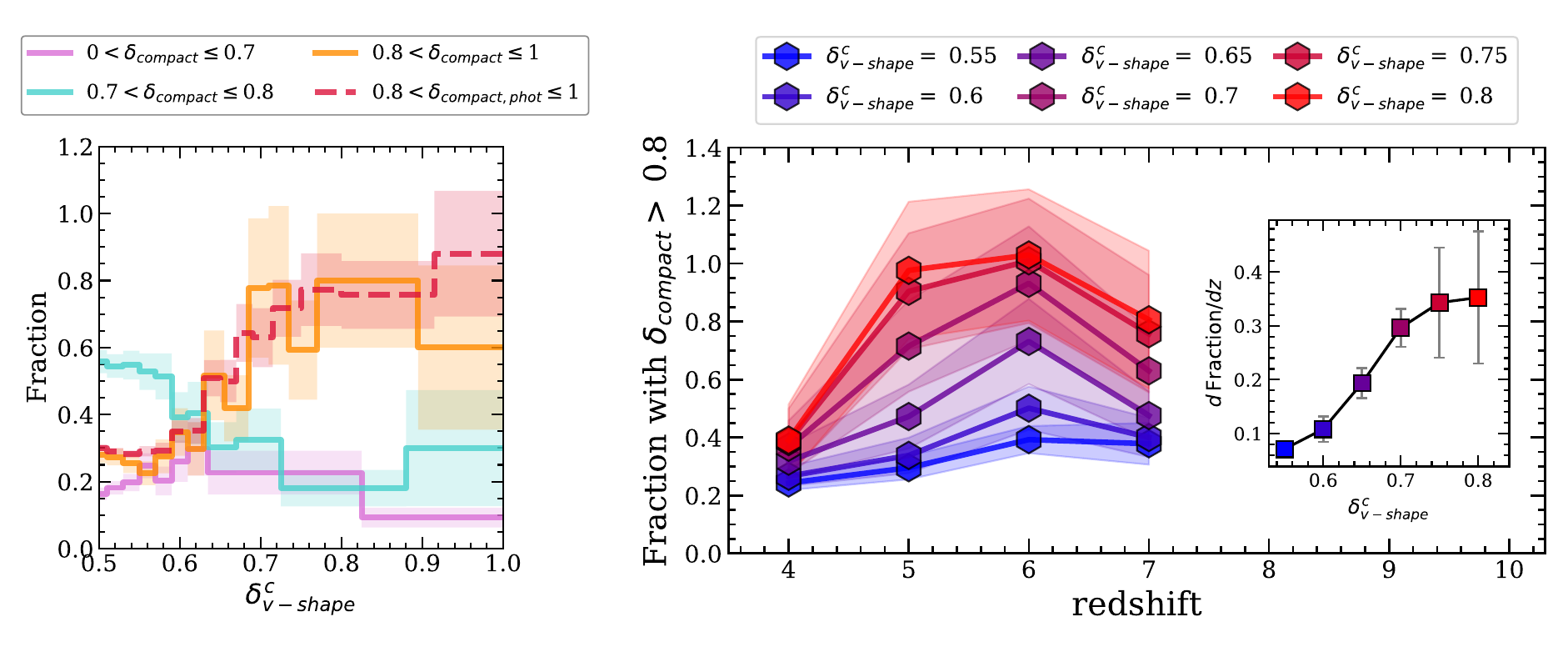}
    \caption{\textit{Left:} Fraction of galaxies (within a given $\delta_\text{compact}$ interval) as a function of the V-shape intensity, $\delta_\text{v-shape}$. The red dashed curve denotes our spectroscopic sample complemented by photometric data. Error bars represent Poisson uncertainties. \textit{Right:} Fraction of compact galaxies with V-shape intensities ranging from $\delta_\text{v-shape}^c=0.55$ to $0.8$, as a function of redshift. The inset shows the gradient of this fraction in the redshift range $4<z<6$ as a function of $\delta_\text{v-shape}$.}
    \label{fig:proportion_compact_v_shape}
\end{figure*}

Based on Fig. \ref{fig:delta_vshape_delta_compact}, we observe that a significant proportion of galaxies with strong $\delta_\text{v-shape}$ values also tend to be compact. This is supported by the left panel of Fig.~\ref{fig:proportion_compact_v_shape}, which illustrates the fraction of compact galaxies as a function of their V-shape intensity, $\delta_\text{v-shape}$. As the SED in galaxies becomes more V-shaped, their morphology tends to be more compact. This result aligns with findings by \citet{Hviding_RUBIES}, who used a smaller sample from the RUBIES survey \citep{deGraaff_RUBIES}, where a large portion ($\approx75-80\%$) of galaxies with V-shaped SEDs were identified as point sources. However, this is the first time such a correlation is shown with high statistics: $\approx 1370$ sources respecting $\delta_\text{v-shape}>0.5$ and $\delta_\text{compact}>0.8$ in our photometric sample, forming the red curve in Fig. \ref{fig:proportion_compact_v_shape}-left, indicating that compactness is an intrinsic property of galaxies with intense V-shaped SEDs.

Moreover, we do not observe any sharp transitions, even around the LRD boundaries ($\delta_\text{v-shape} \approx 0.6$). Instead, the fraction of compact galaxies increases with the intensity of the V-shape. This shows a continuous relationship between the continuum properties ($\delta_\text{v-shape}$) and morphology ($\delta_\text{compact}$) and a global transition toward higher LRD state. At $\delta_\text{v-shape} = 0.6$,  $\approx30\%$ of galaxies exhibit $\delta_\text{compact} > 0.8$, whereas at $\delta_\text{v-shape} = 0.75$, it increases to $\approx 70 \%$ of the galaxies. There is therefore a non-negligible number of galaxies that share the same continuum properties as LRDs but are extended, a population that dominates up to $\delta_\text{v-shape} \approx 0.7$. Beyond this value, compact galaxies dominate, despite large uncertainties due to the lack of statistics. 

We notice that the curve representing the amount of compact galaxies ($\delta_\text{compact} > 0.8$) as a function of the intensity of the V-shape (red curve in Fig. \ref{fig:proportion_compact_v_shape}-left) has a sigmoid shape. By fitting a sigmoid function per interval of redshifts from $z=3$ to $z=8$ (see Appendix \ref{sec:appendix2}), we are also able to probe the amount of compact galaxies at a given V-shape intensity and a given redshift, which is displayed in Fig.~\ref{fig:proportion_compact_v_shape}-right. We recover the typical redshift range for LRDs, between $4 < z < 8$ \citep[e.g.,][]{kocevski_rise_2024, akins_cosmos-web_2024, Ma_counting_LRD_2025}, with a peak around $z \approx 6$. Given the magnitude threshold applied ($m_{\text{H}\alpha}$<27.5 AB mag; see Appendix ~\ref{sec:appendix0}), the decline at the high-redshift end is likely due to incompleteness, as also discussed in \citet{kocevski_rise_2024} and \citet{Pacucci2023}. Additionally, we observe a similar redshift-dependent behavior for all $\delta_\text{v-shape} > 0.5$ values, with an increase in the contribution of compact galaxies as the intensity of the V-shape increases. The steepness of the decline in the $4<z<5$ range strongly depends on $\delta_\text{v-shape}$, with a steeper decline for sources exhibiting a more intense V-shape, as shown by the subplot in the right panel of Fig. \ref{fig:proportion_compact_v_shape}. This suggests a shorter life time or duty cycle of such extreme sources, potentially quickly evolving to more extended, less V-shaped objects \citep{Billand_investigating}.

\section{Evolution of the Balmer break strength}\label{sec:BBK}
Several extreme LRDs exhibit, alongside other properties, a Balmer break strength incompatible with a stellar origin \citep{Ji_black_thunder,Naidu_bhstar,degraaff_a_remarkable,Taylor_2025_capers}. \cite{DeGraaff_BHstar} suggested that the optical emission in LRDs originates from a modified black body, which would form extreme Balmer breaks with a temperature of $T_\text{BB} \approx 5000$\,K.

Therefore, we sought to investigate this alternative property, as the Balmer break strength could serve as a distinguishing feature for the LRD population. To compute the Balmer break strength, we adopt the same definition as described in \citet{degraaff_a_remarkable}, using the ratio of the mean flux density between $[3620, 3720]$\,\AA\ and $[4000, 4100]$\,\AA. We utilize the PRISM spectra available on the DJA, retaining only those that match our photometric catalog (see Sect. \ref{spectroscopic_data}). After excluding measurements with low robustness (S/N $< 2$), we retained $\approx$5000 sources with robust Balmer break measurements. It is important to note that computing $\beta_\text{opt}$ following the method of \citet{kocevski_rise_2024} relies on broadband filters where the bluest data point remains redward of the Balmer break. Consequently, this photometric approach probes redder wavelengths that do not overlap with the spectral range used to calculate the Balmer break strength spectroscopically. In this work, these two quantities are therefore distinct and independent of the method used to compute the optical slope. This contrasts with \citet{barro_from}, where the optical slope is calculated across the Balmer break (from 0.3$\mu$m to 0.9$\mu$m rest-frame), inherently correlating the optical slope and Balmer break strength by construction.

In Fig.~\ref{fig:BBK_delta_compact} we present the Balmer break strength as a function of $\delta_\text{compact}$ across different bins of V-shape intensity: $\delta^c_{\text{v-shape}} < 0.55$, $0.55 < \delta^c_{\text{v-shape}} < 0.65$, and $\delta^c_{\text{v-shape}} > 0.65$. These three bins represent non-V-shaped, intermediate, and intense V-shaped sources, respectively. We then compute the median Balmer break for each $\delta_{\text{compact}}$ bin, ensuring that each bin contains at least 15 sources. For the $0.55 < \delta^c_{\text{v-shape}} < 0.65$ and $\delta^c_{\text{v-shape}} < 0.55$ cases, we require 50 sources per bin to reduce the total number of bins and improve visual clarity. Surprisingly, as we go toward compact V-shape sources, we do not observe a clear trend for them to present a prominent Balmer break. Effectively, the most extreme bin  ($\delta_{\text{compact}} >0.9  \ \& \ \delta^c_{\text{v-shape}} >0.65$), presents a median value of $1.3^{+2.8}_{-0.98}$ similar to the median values of other bins. More generally, no distinct behavior is visible regardless of galaxy morphology or continuum shape. On a subsample of 37 LRDs from \citet{kocevski_rise_2024}, we infer a Balmer break strength of $1.4^{+2.2}_{-0.98}$, in agreement with our values in the LRD regime.

For all the Balmer break strengths computed ($\approx 5000$), the most extremes ones are observed in {CAPERS-LRD-z9} and {MoM-BH*-1}. Given the absence of a correlation between compactness, V-shape intensity, and Balmer break strength, sources with prominent Balmer breaks appear as outliers, even within the usual LRD population. Effectively, sources with a Balmer break strength exceeding 3 account for only $\approx 3\%$ of those exhibiting both a V-shape ($\delta^c_{\text{v-shape}} > 0.6$) and a compact morphology ($\delta_{\text{compact}} > 0.8$). 

This demonstrates that the majority of LRDs do not exhibit a prominent Balmer break, as the median value remains between $1$ and $1.5$ for both extended and compact galaxies, regardless of the V-shape prominence. Coupled with the large scatter observed in compact, V-shaped galaxies, this implies that the population probed in the high LRD regime is not homogeneous.

\section{Behavior of the H$\alpha$+ [N\,\textsc{ii}] complex as a function of $\delta_{\text{v-shape}}$, $\delta_{\text{compact}}$}\label{sec:Halpha_evolution}

\subsection{H$\alpha$ line analysis method}\label{sec:Halpha_evolution_method}

We found that the morphology and continuum are strongly correlated (see Sect.~\ref{subsec:corralation_dvshape_dcompact}), although LRDs also exhibit a third key property: the presence of broad H$\alpha$ and H$\beta$ lines. Given that the H$\alpha$ line is the brightest, we focus exclusively on its behavior. To resolve and decompose the H$\alpha$ complex, we restricted our analysis to the NIRSpec grating, which offers both medium ($R \approx 1000$) and high resolution ($R \approx 3000$). 

To investigate the evolution of the H$\alpha$ line shape, that is, the contribution of the broad component as a function of the continuum ($\delta_{\text{v-shape}}^c$) and compactness ($\delta_{\text{compact}}$), we first divided the galaxies into two groups: compact ($0.8<\delta_{\text{compact}} < 1$) and extended ($0.6<\delta_{\text{compact}} < 0.8$). Then, we stacked the grating spectra per bin of $\delta_{\text{v-shape}}^c$. This approach allowed us to first investigate any global behavior as galaxies approach the LRD state, while also increasing the S/N, particularly in the wings of the H$\alpha$ lines, where the broad component might otherwise remain undetected or partially detected.

\begin{figure}[h!]
    \centering
    \includegraphics[width=9cm]{ 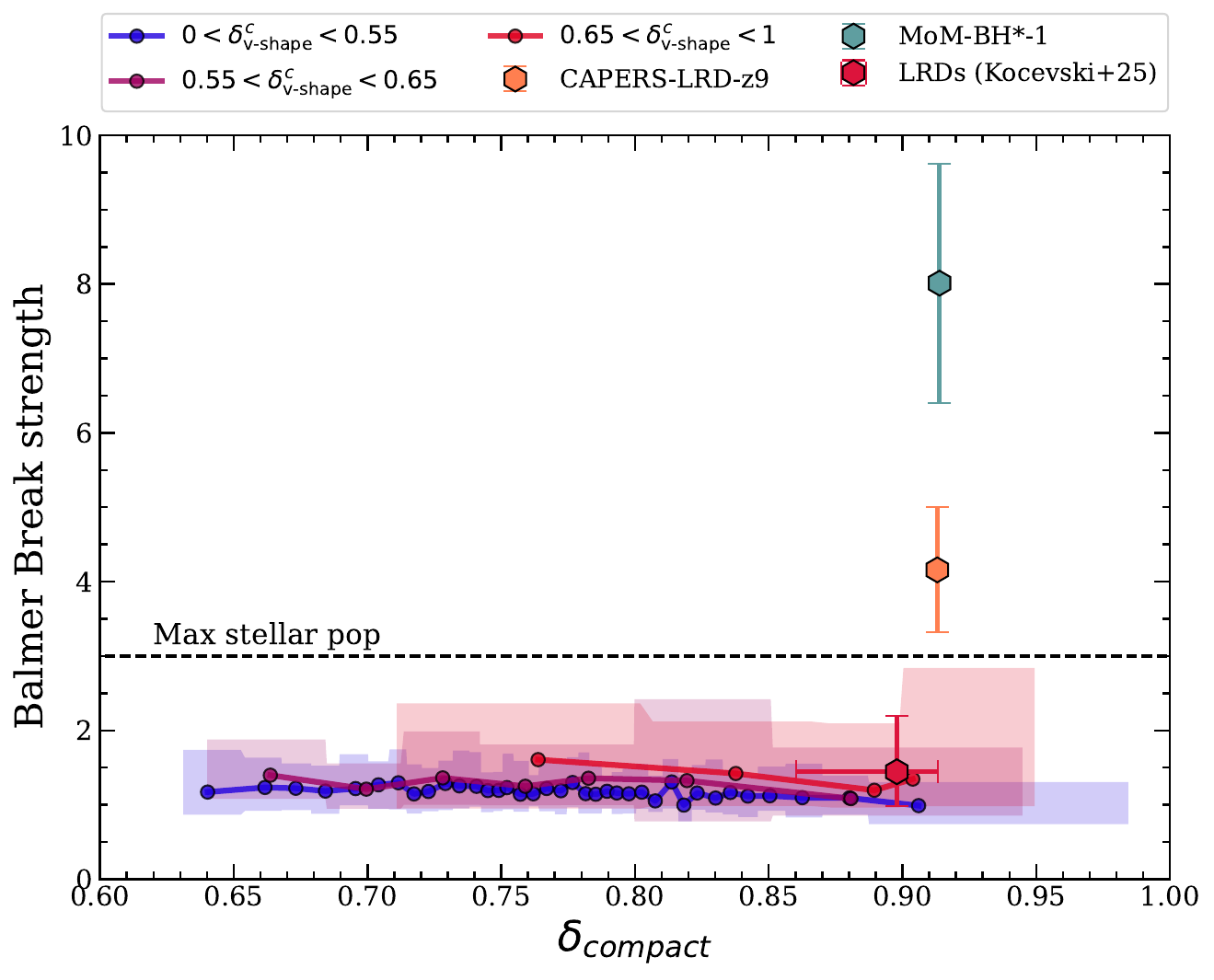}
    \caption{Balmer break strength as a function of $\delta_\text{compact}$ for three bins of $\delta_\text{v-shape}$. Shaded regions correspond to the 1-sigma uncertainties for each bin: the vertical extent represents the 16th and 84th percentiles, while the horizontal width indicates the bin size. The dashed horizontal line denotes the maximum Balmer break strength of stellar origin observed to date \citep{Naidu_bhstar}. The distribution of 37 LRDs from \citet{kocevski_rise_2024} sample is shown as a red hexagon. Cyan and orange hexagon represent Mom-BH*-1 and CAPERS-LRD-z9 from \citet{Naidu_bhstar} and \citet{Taylor_2025_capers}, respectively. Only robust measurements with $S/N > 2$ are included in this figure.}
    \label{fig:BBK_delta_compact}
\end{figure}

To achieve this, we selected only the grating spectra with confirmed spectroscopic redshifts. Each spectrum is then shifted to the rest-frame. To prevent artificial broadening during stacking, arising from uncertainties in the spectroscopic redshift, we refine the redshift by aligning the peak emission of the H$\alpha$ complex, these corrections remain within a few spectral bins (typically < 5). Then, each line is normalized to 1, and all spectra are resampled onto a uniform grid. After a visual inspection, we excluded nine sources from the compact group due to the H$\alpha$ line being near detector edges. This exclusion ensures the highest data quality, especially given the limited statistics of the compact group. Since the majority of the spectra were observed with the G395M filter (see Table~\ref{tab:compact_bins}), we resampled all spectra onto a grid with a resolution of $0.0018/(1+z)$\,$\mu$m per bin, corresponding to the maximum dispersion of G395M. Finally, we compute the median to obtain the median stack in each bin of $\delta_{\text{compact}}$ and $\delta_{\text{v-shape}}^c$. To estimate the uncertainty corresponding to the median, we employed a Monte Carlo (MC) approach. Each individual spectrum was re-simulated 5000 times by incorporating Gaussian noise, where the noise amplitude was determined by the individual error associated with each spectrum. For each simulation, we computed the median spectrum from the noisy spectra. The standard deviation of the resulting distribution of these median spectra is then used as the uncertainty for the median spectrum. To maximize the S/N, we defined our bins of $\delta_{\text{v-shape}}$ with a minimum of 16,000 seconds of effective exposure time. The detailed configuration of these bins is presented in Table~\ref{tab:compact_bins}. Our final spectroscopic sample include $\approx$340 spectra for compact galaxies, and $\approx$886 for extended galaxies, with a V-shape intensity between $\delta_{\text{v-shape}}$=0.6 and 1.

Once the median spectra and their associated uncertainties are constructed, we employ \texttt{BADASS v10.2.0} \citep{BADASS} to decompose the H$\alpha$ complex. Originally designed for analyzing spectra from the Sloan Digital Sky Survey (SDSS), this pipeline is highly adaptable and suitable for JWST spectra. It utilizes a Bayesian approach for all parameters, effectively mitigating issues related to local convergence. The decomposition of the H$\alpha$ complex is performed as follows:

\begin{itemize}
    \item \text{Narrow H$\alpha$ line:}
    This component is fitted with a Gaussian profile, allowing for three free parameters: amplitude (\texttt{AMP}), width ($\sigma$), and velocity offset (\texttt{voff}).

    \item \text{Broad H$\alpha$ line:}
    This component is also fitted with a Gaussian profile, but with only two free parameters:  amplitude (\texttt{AMP}) and width ($\sigma$). The velocity offset (\texttt{voff}) is set to be the same as that of the narrow H$\alpha$ line.

    \item \text{[N\,\textsc{ii}] doublet ([N\,\textsc{ii}]$\lambda$6585 and [N\,\textsc{ii}]$\lambda$6549):}
    Both lines are fitted with Gaussian profiles, but with only one free parameter: amplitude (\texttt{AMP}) of the [N\,\textsc{ii}]$\lambda$6585 line.

    The width ($\sigma$) and the velocity offset (\texttt{voff}) of both lines are constrained to be the same as those of the narrow H$\alpha$ line. Additionally, the flux of the [N\,\textsc{ii}]\,$\lambda$6549 line is constrained by atomic physics: by fixing the line width, we can constrain the amplitude such that \texttt{AMP$_{\text{[NII]$\lambda$6549}}$} = \texttt{AMP$_{\text{[NII]$\lambda$6585}}$} $\times$ 0.341 \citep{Osterbrock_lines}.

\end{itemize}

The model used to describe the H$\alpha$ complex includes a total of six free parameters. Additionally, a power law component is incorporated to account for any potential continuum contribution, introducing two further parameters. To ensure physically meaningful results, priors are imposed on the model parameters: the widths of the narrow lines are restricted to the range $0 < \sigma_\text{narrow} < 500$ km s$^{-1}$,
while the width of the broad line is confined to $500 < \sigma_\text{broad} < 1500$ km s$^{-1}$. Furthermore, the velocity offsets of all lines are constrained within the interval $-800 < \texttt{voff} < 800$ km s$^{-1}$. To evaluate the presence of the broad component, we re-fit all the spectra without including a broad component and compute the Bayesian information criterion (BIC) parameter. We considered that a broad component is effectively present and necessary when $\Delta \text{BIC} = \text{BIC}_\text{narrow} - \text{BIC}_\text{narrow+broad}$ exceeds 5 (see Appendix \ref{sec:appendix3}) and the S/N of the broad component is above 5. A similar criterion has been employed in other broad line detection works \citep[e.g.,][]{Juodzbalis_bh_mass, Taylor_broad_line_agn_2024, Brooks_stack_spectra}.

 \begin{figure}[h!]
    \centering
    \includegraphics[width=8cm]{ 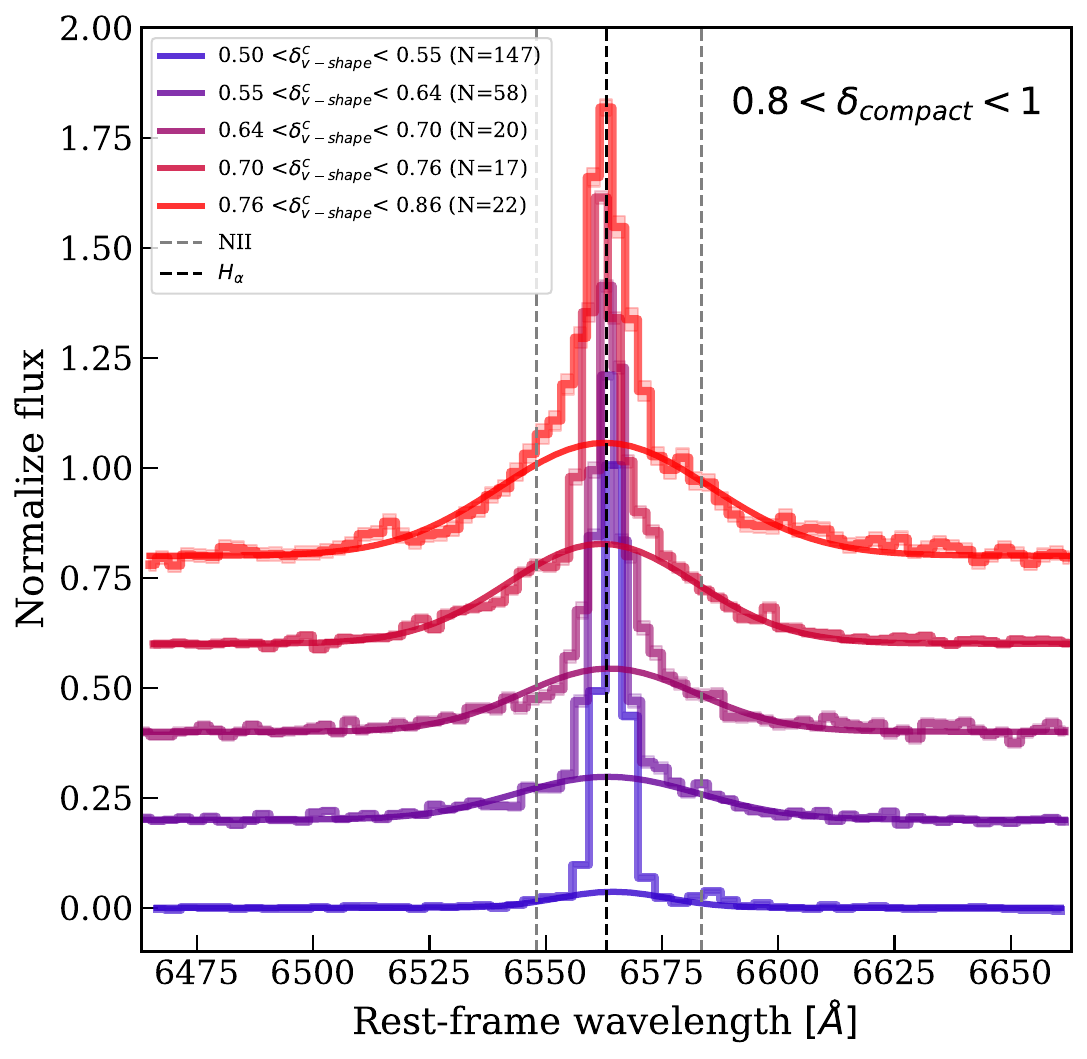}
    \caption{ Stacked spectra of the H$\alpha$ complex along with the result of the broad H$\alpha$ fit. Each spectrum is offset for clarity. The legend specify each bin of $\delta_\text{v-shape}$ and the number of spectra used (always having an effective exposure time above 16,000s), details on the bins can be seen on Table \ref{tab:compact_bins}.
}
    \label{fig:line_profile}
\end{figure}

\subsection{Result of the H$\alpha$ decomposition}

In Fig.~\ref{fig:line_profile}, we show the stacked spectra for each bin of compact galaxies ($  0.8 <\delta_{\text{compact}} < 1$) with the results of the broad-line fit. For each bin of $\delta_{\text{v-shape}}^c$, the presence of the broad component is strongly favored, with a minimum S/N of 9 for the broad component and with $\Delta \text{BIC} > 10$ systematically (see Table~\ref{tab:compact_bins} and Appendix~\ref{sec:appendix3}). All the fits with the residuals can be seen in Appendix \ref{sec:appendix4}. In Fig.~\ref{fig:broad_h_bopt} we display the contribution of the broad component to the total H$\alpha$ luminosity as a function of $\delta_{\text{v-shape}}^c$. A clear trend emerges for compact galaxies. Based on visual only, we fit the relation with a power law of the form $y = A - \left(\frac{x}{x_\text{median}}\right)^{\gamma}$. By normalizing to the median, we helped the convergence and avoid degeneracy between the parameters. We inferred the two following relations,

\begin{equation}
\frac{L(\text{H}\alpha_{\text{broad}})}{L(\text{H}\alpha_{\text{total}})} = 1.53^{+0.03}_{-0.03} - \left(\frac{\delta_{\text{v-shape}}}{0.67}\right)^{-1.25^{+0.16}_{-0.13}} \quad (\delta_{\text{compact}} > 0.8).
\end{equation}

\noindent We employed a Markov chain Monte Carlo approach using the \texttt{emcee} package to fit the parameters and infer the uncertainties. For extended galaxies, we also detected a correlation that is, however, less strong than for compact galaxies. Due to the lack of S/N of the broad component in extended galaxies that limit the number of bins, we cannot conclude whether extended and compact galaxies follow the same relation, but the correlation is still similar for both, indicating a physical connection between the origins of the broad line in extended and compact galaxies. Our results demonstrate a clear three-way correlation: as the V-shape becomes more prominent and the galaxy becomes more compact, the contribution of the broad component increases systematically. No discontinuities or bi-modalities are observed in the behavior of H$\alpha$, which reinforces the continuity in the observed properties identified in Sect. \ref{subsec:corralation_dvshape_dcompact} between the continuum and the compactness.

\subsection{Dichotomy of the [N\,\textsc{ii}]}\label{sec:NII}

The [N\,\textsc{ii}] line is known to be undetected in LRDs \citep{matthee_little_2024, Taylor_broad_line_agn_2024, Hviding_RUBIES} and, hence, poorly studied to date, although it could be used as a feature to compare LRDs with the general population. In Fig. \ref{fig:NII}, we display the ratio of the luminosity of the [N\,\textsc{ii}]$\lambda$6585 luminosity over the narrow H$\alpha$ line as a function of $\delta^c_\text{v-shape}$. A clear dichotomy apparent between compact and extended galaxies. As expected, we recovered the fact that LRDs ($\delta^c_\text{v-shape}$>0.6, $\delta_\text{compact}$>0.8 ) do not present any [N\,\textsc{ii}] components, but more generally, compact galaxies present a global behavior with an extremely low contribution of [N\,\textsc{ii}]. However, it is not the case for extended galaxies, which present a stronger contribution of [N\,\textsc{ii}] as $\delta^c_\text{v-shape}$ increases.

\begin{figure}[h!]
    \centering
    \includegraphics[width=8cm]{ 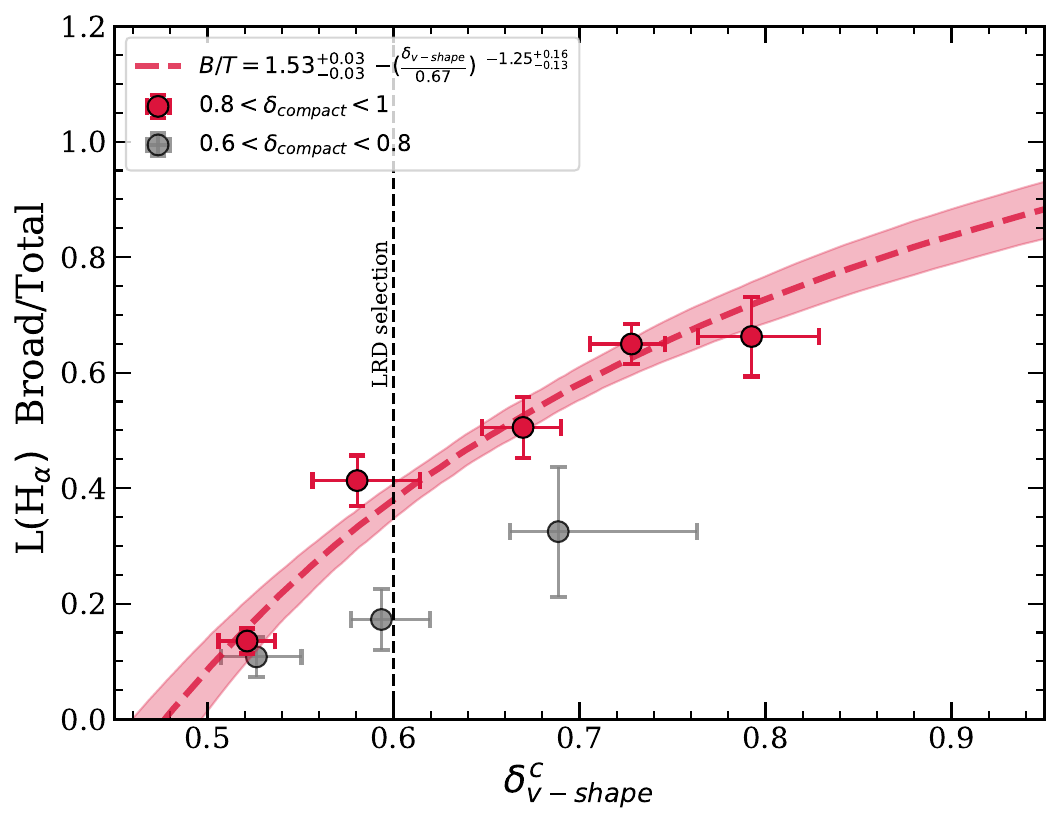}
    \caption{Contribution of the broad H$\alpha$ component as a function of the $\delta_\text{v-shape}^c$ parameter, corrected for the H$\alpha$ boosting effect. The relation for compact galaxies ($0.6<\delta_\text{compact} <1$) is shown in red, while the relation for extended galaxies ($0.6<\delta_\text{compact} < 0.8$) is displayed in gray. }
    \label{fig:broad_h_bopt}
\end{figure}

The absence of [N\,\textsc{ii}] in compact galaxies, aside from the LRDs, could indicate either very low metallicity or the presence of nitrogen in a highly ionized state, potentially ionized by a powerful central engine. Given that compact galaxies tend to host a less massive stellar component \citep{guia_sizes_2024} and that less massive galaxies are typically metal-poor \citep{Gallazzi_metallicity}, it is plausible that the metallicity is too low for the [N\,\textsc{ii}] to be detected even in stacked studies. High-ionization lines have been observed in some LRDs, such as N\,\textsc{iv}$\lambda\lambda 1483$, C\,\textsc{iv}, and [Ne\,\textsc{iii}]$\lambda$3869 \citep{Tripodi_LRD_Z, Lambrides_iron_ionized, Jones_dormant_LRD}, showing that certain complex elements are present in an ionized state. In some LRDs, the metallicity has been inferred and is systematically low, with  $Z < 0.1\,Z_\odot$ \citep[e.g.,][]{Tripodi_LRD_Z, Jones_dormant_LRD, maiolino_pristine, Ji_black_thunder, torralba_Z_LRD}. Moreover, the strong evolution of the broad H$\alpha$ component with $\delta_{\text{v-shape}}$, contrasted with the lack of evolution in the nitrogen emission for compact galaxies, imply that there is no clear link between the origin of the broad line and the absence of [N\,\textsc{ii}]. However, such a link would be expected if the nitrogen was ionized by the AGN, with the broad line originated from  the same AGN, as for instance interpreted in the BPT diagram \citep{Baldwin_BPT}. This observation also supports the hypothesis of metal-poor compact sources rather than AGN-driven [N\,\textsc{ii}] emission. This aligns with the findings of \citet{kocevski_hidden_2023} and \citet{harikane_jwstnirspec_2023-1}, which highlight the limitations of the BPT diagram at high redshifts for diagnosing AGNs in metal-poor environments. The behavior observed is also consistent with the results for extended galaxies, which systematically exhibit a more significant [N\,\textsc{ii}] contribution. The stellar component in extended galaxies naturally increases with metallicity, allowing for the detection of [N\,\textsc{ii}]. If the broad line in extended galaxies is interpreted as a proxy for the central BH mass, it suggests that the more massive the central BH, the more prominent the nitrogen emission. This can be interpreted similarly to the behavior of regular AGNs or through the BPT diagram: the more powerful the central AGN, the more likely we are to observe the [N\,\textsc{ii}] line. \\

\begin{table*}[t]
\centering
\caption{Description of the $\delta_{\text{v-shape}}$ bins for compact sources ($\delta_{\text{compact}}>0.8$).}
\label{tab:compact_bins}
\setlength{\tabcolsep}{5pt} 
\begin{tabular}{l c c c c c c c c}
\hline \hline
\noalign{\smallskip}
Bin range & $\delta_{\text{v-shape}}$ & N$_\text{spec}$ & $z$ & F444W & $t_{\text{exp}}$\tablefootmark{a} & Grating\tablefootmark{b} & S/N & FWHM \\
$\delta_{\text{v-shape}}$ & median & & median & [mag] & [$10^3$s] & G395M/G395H/G235M & broad & [km/s] \\
\noalign{\smallskip}
\hline
\noalign{\smallskip}
[0.5, 0.55]  & $0.520^{+0.016}_{-0.013}$  & 147 & $4.10 \pm 1.11$ & $26.28 \pm 1.05$ & 153.2 & 96 / 17 / 34   & 9.5  & $1261\pm 56$\\
]0.55, 0.64] & $0.578^{+0.035}_{-0.02}$  & 58 & $5.28 \pm 1.29$ & $25.88 \pm 1.52$ & 59.4 & 46 / 7 / 5   & 12.9 & $2120 \pm 84$\\
]0.64, 0.70] & $0.655^{+0.025}_{-0.017}$  & 20 & $5.82 \pm 1.04$ & $26.08 \pm 1.12$ & 20.9 & 18 / 2 / 0   & 13.0 & $2017 \pm 81$\\
]0.70, 0.76] & $0.729^{+0.019}_{-0.022}$  & 17 & $5.50 \pm 0.76$ & $25.46 \pm 0.91$ & 16.3 & 15 / 0 / 2   & 24.4 & $2180 \pm 40$\\
]0.76, 0.86] & $0.793^{+0.034}_{-0.029}$  & 22 & $5.55 \pm 1.14$ & $25.74 \pm 1.18$ & 21.0 & 20 / 2 / 0   & 16.0 & $2444\pm 86$\\
\noalign{\smallskip}
\hline
\end{tabular}
\tablefoot{
\tablefoottext{a}{Effective exposure time in $10^3$s.}
\tablefoottext{b}{Number of spectra per grating configuration.}
}
\end{table*}

To further investigate the metallicity, we computed the O32 and Ne3O2 parameters. Following the methodology presented in Sect.~\ref{sec:Halpha_evolution_method}, we stacked the PRISM spectra as a function of $\delta_{\text{v-shape}}$, restricting the selection to sources with $4<z<7$ to minimize to probe to different epoch, as metallicity is known to evolve with redshift \citep[e.g.,][]{Maiolino_metallicity_z, Sanders_metallicity}. Another parameter widely used is the R23 parameter, define as $R_{23} = ([O_\textsc{iii}] \lambda4963,5007 + [O_\textsc{ii}] \lambda3727) / H\beta$. Since R23 is normalized by H$\beta$, which likely presents a non negligible contribution from the central engine in the LRD context, we excluded it from our work. 

Here, O32 is defined as  $O_{32} = [O_\textsc{iii}]\lambda5007 / [O_\textsc{ii}]\lambda \lambda3727,29$ and Ne3O2 as Ne3O2=$[Ne_\textsc{iii}]\lambda3870 / [O_\textsc{ii}]\lambda3728$. Once these are computed, we can translate both parameters into metallicity ($12 + \log(\text{O/H})$), using the calibration inferred for high-redshift galaxies ($2 < z < 9$) by \citet{Sanders_metallicity}. In Fig. \ref{fig:metelicities} we display the mean metallicity inferred from the O32 and Ne3O2. We recover the dichotomy we observed on the [N\,\textsc{ii}]: compact galaxies tend to be metal poor ($Z<0.1Z_\odot$) compared to extended galaxies ($Z>0.1Z_\odot$). This implies that LRDs are metal-poor primarily because of their compactness, rather than their classification as LRDs. Within the broader population of compact objects, however, we tend to identify a further evolution toward lower metallicities as the V-shape intensity increases, potentially probing lower mass host galaxies compared to other compact systems.

The emission lines used to derive metallicities in this work can be affected by high electron densities ($n_e$), which may trigger collisional de-excitation in dense environments. However, several features suggest that the [O\,\textsc{iii}] and UV emission originate from the host galaxy rather than the dense central engine. These include the observed correlation between [O\,\textsc{iii}] luminosity and $M_{\text{UV}}$ \citep{DeGraaff_BHstar}, the lack of UV variability \citep{furtak_investigating_2025}, and the presence of extended UV emission \citep{Rinaldi_beyond, chen_host_2024, Baggen_connecting}. Consequently, the low metallicities inferred for LRDs and compact galaxies favor a low-mass host scenario \citep[e.g.,][]{DeGraaff_BHstar, sun_bhstar_host}. This is consistent with a young stellar component, which has been previously proposed to explain the rest-frame UV emission in LRDs \citep[e.g.,][]{Taylor_2025_capers, Delvecchio_stack, torralba_Z_LRD, barro_from}.

\begin{figure}[h!]
    \centering
    \includegraphics[width=6cm]{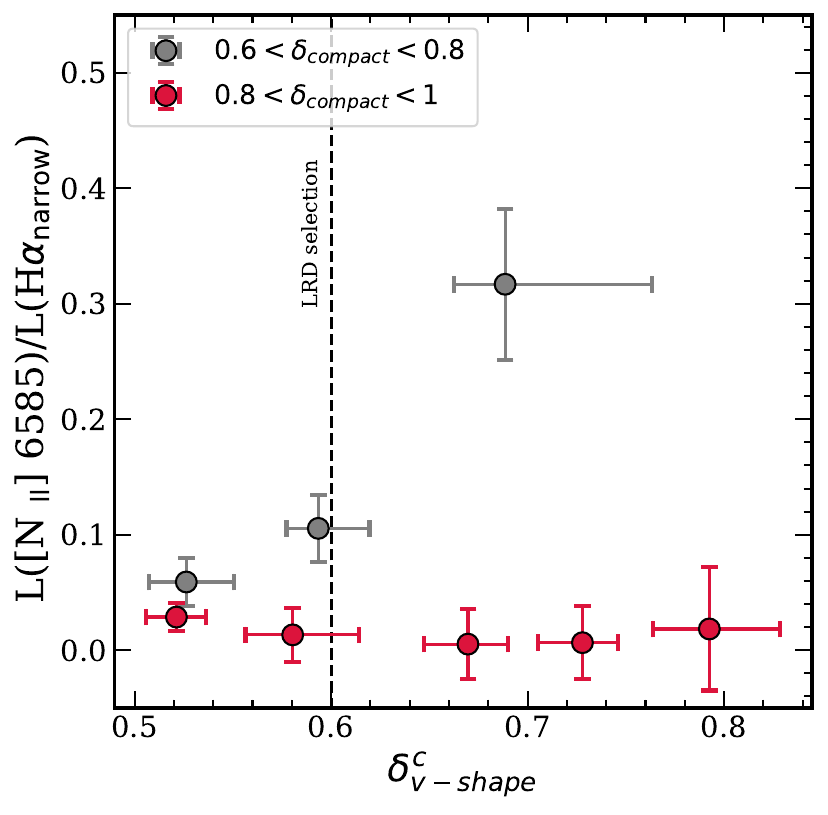}
    \caption{ Contribution of the [N\textsc{ii 6585}] to the H$\alpha$ narrow line, as a function of $\delta_{\text{v-shape}}^c$, for compact (in red) and extended galaxies (in gray).
}
    \label{fig:NII}
\end{figure}

\section{Balmer decrement}

The Balmer decrement observed in LRDs is surprisingly high, reaching values comparable to those found in heavily obscured dusty galaxies: H$\alpha/$H$\beta \approx 10$ (e.g., \citealp{DeGraaff_BHstar}) and even $\approx 16$ in the BH$^*$ interpretation \citep{sun_bhstar_host}. Following the procedure described in Sect. \ref{sec:NII}, we stacked the PRISM data and performed the line fitting using \texttt{BADASS}, focusing on the $0.40$--$0.70~\mu$m rest-frame range to capture both H$\beta$ and H$\alpha$. We computed the observed Balmer decrement as a function of $\delta_{\text{v-shape}}^c$ and $\delta_{\text{compact}}$, for sources between $4<z<7$, as shown in Fig. \ref{fig:balmer_decrement}. Our results confirm that in the LRD regime, the Balmer decrement is significantly elevated ($\approx 10$). More generally, we find that the Balmer decrement correlates strongly with $\delta_{\text{v-shape}}^c$, regardless of the galaxy's compactness. This suggest that both the V-shape intensity and the Balmer decrement share common physical origins, within and outside the LRD regime. In field galaxies, the fact that the Balmer decrement increases when the continuum becomes redder suggests that this common behavior is produced by dust attenuation. Since LRDs follow a similar trend, this suggests that this common behavior also originates from dust attenuation.

Under the assumption of a dust attenuation common origin, the Balmer decrement can be predicted for a given continuum shape, characterized here by $\delta_{\text{v-shape}}^c$, and a given attenuation law. By measuring the median optical slope ($\beta_{\text{opt}}$) within each $\delta_{\text{v-shape}}^c$ bin and adopting two different attenuation laws \citep{Calzetti_2000,Gordon_2003}, we can map the continuum shape to a visual attenuation ($A_V$), and subsequently to a predicted Balmer decrement. Since the relative attenuation between gas and stars is uncertain at high redshift \citep[see, e.g.,][]{Shivaei_attenuation_z}, we adopt a $\frac{E(B-V)_\text{star}}{E(B-V)_\text{gas}}$ factor of 0.44 for the \citet{Calzetti_2000} law, and a value of 0.74 for the Small Magellanic Cloud (SMC) law, following \citet{Theios_attenuation_EBV}. This illustrates the range of possibilities and the general uncertainties associated with this comparison. This prediction also requires the assumption of an intrinsic optical slope, $\beta_{\text{opt,int}}$. In the case of a stellar-dominated system, we simulated SEDs using \texttt{CIGALE} \citep{cigale} and inferred $\beta_{\text{opt,int}}$ values ranging from $-2.3$ to $-1.8$ \footnote{Assumed stellar ages range from $0.1$ to $1.5$~Gyr, where $\approx 1.5$~Gyr corresponds to the age of the universe at $z \approx 4$ }. We note that the intrinsic optical slope proposed in the recent AGNs model by \citet{madau_maiolino_LRDs} is also $\approx -2$. This implies that whether the intrinsic flux originates from a stellar population or an AGN; the two would exhibit similar observed behavior.  \\

\begin{figure}[h!]
    \centering
    \includegraphics[width=6cm]{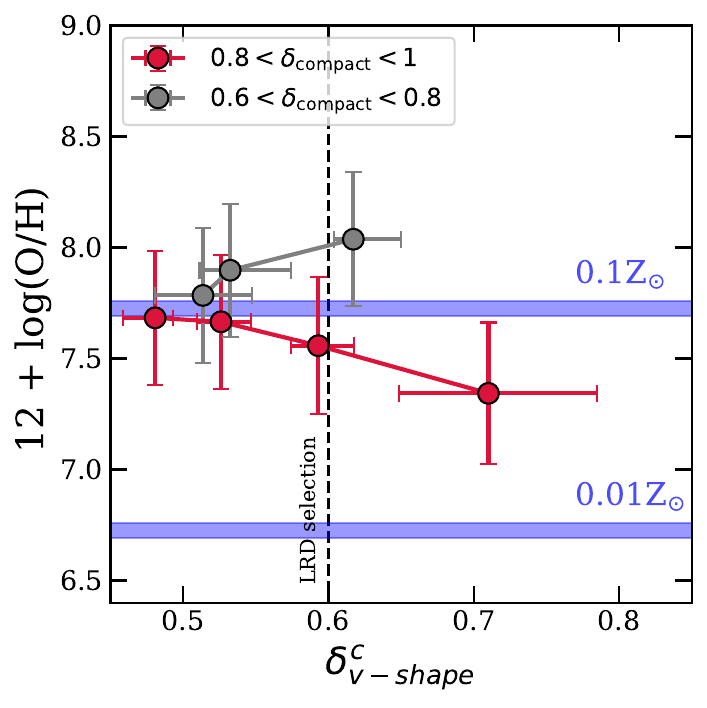}
    \caption{Metallicity inferred with the O32 and Ne3O2 parameters between 4<z<7, as a function of $\delta_{\text{v-shape}}$. Red (gray) circles denote compact (extended) galaxies.
}
    \label{fig:metelicities}
\end{figure}

\begin{figure}[h!]
    \centering
    \includegraphics[width=8.5cm]{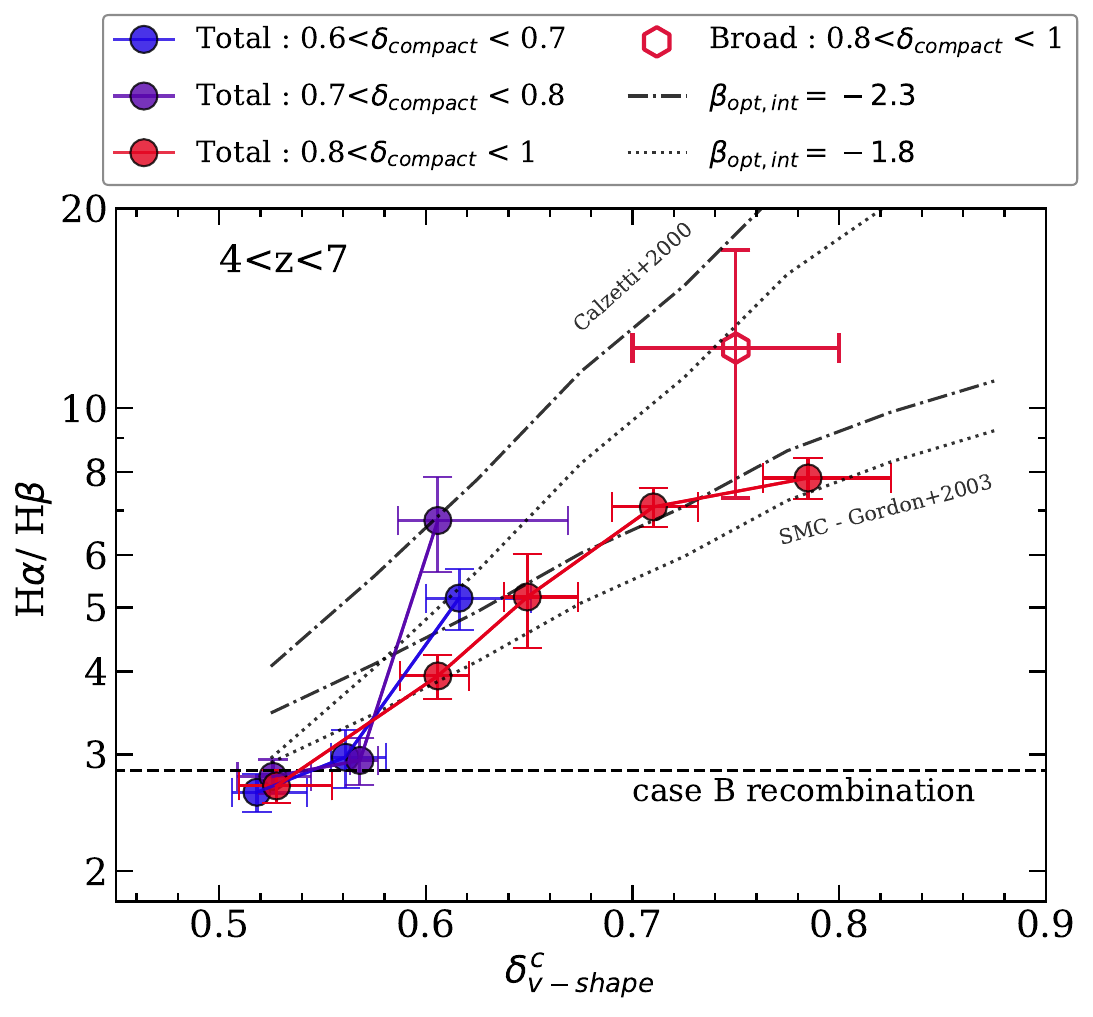}
    \caption{Observed Balmer decrement as a function of  $\delta_{\text{v-shape}}^c$, for different bins of compactness $\delta_{\text{compact}}$. The dotted line represent the Case B recombination and the other lines (dotted and dash-dotted) the Balmer decrement expected from dust attenuation.}
    \label{fig:balmer_decrement}
\end{figure}

As shown, the observed increase in H$\alpha/$H$\beta$ with $\delta_{\text{v-shape}}^c$ is consistent with a dust-attenuation interpretation across the entire sample, including the LRD regime. More specifically, the majority of the data points appear consistent with an SMC attenuation law \citep{Gordon_2003}, except for the extended, v-shaped sources that tend to follow this time the \citet{Calzetti_2000} attenuation law. This is in line with our result of Fig. \ref{fig:metelicities}, in which more extended galaxies tend to be more metal rich and are expected to follow the \citet{Calzetti_2000} law. 

Adopting the SMC attenuation law, we derive $A_V$ from H$\alpha /$H$\beta$ in the LRD regime (last 3 red points in Fig. \ref{fig:balmer_decrement}). Following the methodology of \citet{casey_dust_2024} and using the median $R_e$ in each bin (provided by \citealt{Genin_DJA} on the DJA), we convert $A_V$ into an estimated dust mass. This provides a robust constraint on the maximal dust content of LRDs, which is estimated to be $M_\text{dust max, LRDs} \approx 3 \times 10^{4}~M_\odot$. This is consistent with the results of \citet{casey_dust_2024}, who used an $A_V$ derived from SED fitting. Since $M_\text{dust} \propto R_e^2$, we interpret this value as an upper limit, as we assumed the overall object to be impacted by dust: the $R_e$ used is probing the overall object and is overestimating the spatial attenuated region if the latter is confined toward the central engine for instance, which as been suggested in \citet{Nikopoulos_evidence}. This dust mass could therefore be interpreted under a purely stellar assumption. Moreover, contributions to the rest-frame optical flux in LRDs beyond the stellar continuum such as AGN activity \citep[e.g.,][]{barro_from} or as recently suggested, significant nebular emission \citep{Sneppen_nebular}, would contribute to lowering this inferred mass as also discussed hereafter.
 Nevertheless, this implies that even a modest amount of dust, if present, can account for both the observed Balmer decrements and the red rest-frame colors of LRDs.

The Balmer decrement in most LRDs could effectively be dominated by the broad component; however, with our current data, we managed to infer only one robust value for the Balmer decrement of the broad component in LRDs,  (H$\alpha /$H$\beta$)$_\text{broad}= 16 \pm 5$, visible as a red open hexagon on \ref{fig:balmer_decrement}, reinforcing its link to the central engine. \citet{DeGraaff_BHstar} and \citet{sun_bhstar_host} suggested that the Balmer decrement might reflect the physical conditions of a dense gaseous medium, where scattering and collisional de-excitation drive the high H$\alpha/$H$\beta$ ratios. The origin of the decrement is therefore likely due to either dense-gas-driven processes or dust attenuation, with the latter potentially being linked to a dusty BLR, as suggested by \citet{madau_maiolino_LRDs}. We note that the observed increases of the broad component in Sect. \ref{sec:Halpha_evolution}, taken together with the observed trend of the Balmer decrement in Fig. \ref{fig:balmer_decrement}, suggest first that the central engine contribute further at hight $\delta_\text{v-shape}$ values, and then that it is compatible with dust attenuation, supporting the dusty BLR interpretation. 

To further investigate the origin of the Balmer decrement in LRDs, we examined higher order Balmer transitions in our stacked PRISM spectra. We specifically focus on H$\delta$, as H$\gamma$ is frequently blended with the $[OIII]\lambda4363$ auroral emission. Figure~\ref{fig:balmer_lines_ratio} shows the observed Balmer decrement (H$\alpha$/H$\beta$) as a function of the H$\delta$/H$\alpha$ ratio for the compact sources in our sample ($0.8 < \delta_{\text{compact}} < 1$), color-coded by the mean value of $\delta_{\text{v-shape}}^c$ in each bin, a similar figure is present in \citet{Nikopoulos_evidence}. Unlike the analysis presented in Fig.~\ref{fig:balmer_decrement}, which partially depends on the assumed intrinsic optical slope ($\beta_{\text{opt,int}}$) and, thus, on a stellar-dominated and AGN continuum in our case, these line ratios are independent of the intrinsic continuum shape and rely solely on Case~B recombination theory. Assuming a \citet{Calzetti_2000} extinction law, we predict the expected ratios and include a $5\%$ scatter to account for intrinsic variations in Case~B conditions (see Fig.~\ref{fig:balmer_lines_ratio}). The mean observed Balmer ratios in LRDs is consistent with dust extinction.

However, when examining individual data points, the one exhibiting the most pronounced v-shape shows a slight deviation from the attenuated Case B prediction, which could be interpreted as a process other than attenuation being at play, such as collisional de-excitation. We note, though, that the equivalent width (EW) of the H$\delta$ emission line in this specific bin is particularly low, making it sensitive to any underlying absorptions feature, as an absorber in the line of sight or stellar Balmer absorption lines. Indeed, for the latter, several arguments support the presence of a young stellar population in LRDs, as discussed in Sect.~\ref{sec:NII}. Such a population, dominated by O, B, and A-type stars, would naturally exhibit intrinsic Balmer absorption features. We notice however that the absorption lines already detected in LRDs \citep[see, e.g.,][]{kocevski_rise_2024, barro_from} are unlikely of stellar origin, as they have been found to be broad and offset in a few sources \citep{Davis_ocean}, but more related to an absorber in the line of sight. In any case, they still contribute to lowering the observed EW of the Balmer lines, which is not corrected in the observed values.

To account for this, we applied illustrative EW corrections of approximately 2, 5, and 6\,\AA\ to the H$\alpha$, H$\beta$, and H$\delta$ lines, respectively, based on findings at $z \approx 2$ by \citet{seille_balmer}. These corrections compensate for potential underlying stellar absorption \citep[see also][]{Hopkins_balmer_absorption, Groves_balmer_dec_correction} or any other absorbers along the line of sight. When applied (shown as an open hexagon in Fig.~\ref{fig:balmer_lines_ratio}), these corrections shift the observed ratios toward the \citet{Calzetti_2000} extinction law. Although this correction might be overestimated, it illustrates the dynamical range of the uncertainties in our current data. It also highlights that the deviation from the Case~B prediction for the reddest point is primarily driven by these uncertainties. Regardless of whether this correction is applied, the mean LRD ratios, which include the three most V-shaped points, remain consistent with dust-attenuated Case~B recombination. We infer mean observed ratios of $(\text{H}\alpha/\text{H}\beta)_{\text{LRDs}} = 6.7 \pm 1.1$ and $(\text{H}\delta/\text{H}\alpha)_{\text{LRDs}} = 0.02 \pm 0.01$. We therefore conclude that the Balmer line ratios, and consequently the Balmer decrements, in most LRDs are consistent with dust extinction, although we cannot rule out secondary contributions that remain indistinguishable with the present data.

Since we observe a similar correlation between non-LRD and LRD sources, as well as a continuous transition as sources approach high LRD states, and given that the Balmer decrement is driven by dust in the non-LRD regime and consistent with attenuation in the LRD regime, this suggests that dust could be responsible for the observed Balmer decrement in LRDs. As showed by \citet{casey2025upperlimit106modot}, current ALMA upper limit on the dust mass in LRDs is $10^6$M$_\odot$, which is more than 1 dex above our M$_\text{dust max, LRDs}$. This imply that the presence of dust cannot be excluded.

Moreover, within the BH$^*$ framework, the predicted Balmer decrement is highly sensitive to the gas column density, $N_\text{HI,2s}$ \citep[see Fig. 5 in][]{Chang_lines_bhstar}. If the entire LRD population were effectively composed of BH$^*$ with $N_\text{HI,2s}>10^{14}$cm$^{-2}$, a clear discontinuity with the galaxy field would be expected, which was not observed here.

\section{Discussion }\label{sec:discussion}

In this section, we synthesize our findings and discuss their interpretation within the context of LRDs. We first establish that the continuum shape of all galaxies is strongly correlated with morphology, as galaxies exhibiting a pronounced V-shape tend to be more compact (see Fig. \ref{fig:proportion_compact_v_shape}-left). Moreover, we observe a smooth progression toward the LRD state, with no evidence of a sharp discontinuity around standard LRD selection criteria.

We further demonstrate that broad-line intensity is also strongly correlated with the shape of the continuum as probed by the so-called V-shapeness measured by our parameter $\delta_{\text{v-shape}}$, indicating a clear connection between these three properties (see Fig. \ref{fig:broad_h_bopt}). While this trend is more pronounced in compact sources, it persists among extended sources as well. Although compact, V-shaped sources are more likely to exhibit non-stellar Balmer breaks, the majority still display typical Balmer break strengths, with values close to unity (see Fig. \ref{fig:BBK_delta_compact}). Additionally, compact sources generally lack a significant [N\,\textsc{ii}] component, whereas [N\,\textsc{ii}] emission tends to be more prominent in extended galaxies (see Fig. \ref{fig:NII}). We interpret this as a direct consequence of lower metallicity in compact galaxies, an hypothesis further supported by metallicity estimates inferred from the O32 and Ne3O2 parameters. The Balmer decrement scales strongly with the intensity of the V-shape, as more pronounced V-shaped sources exhibit a higher H$\alpha$/H$\beta$ ratio. Furthermore, extended galaxies show a higher decrement at a given V-shape intensity. In particular, all these trends are continuous, with no observed discontinuities near the boundaries of LRD selections. This suggests a gradual progression toward the LRD state, rather than a discrete "on-off" transition that would define LRDs as a strictly distinct population undergoing different physical processes.

The first key property we inferred in this work is that the origin of broad-line emission is intrinsically linked to the optical continuum, as evidenced by their strong correlation across all observed morphologies. In the framework of the BH$^*$ interpretation \citep{Naidu_bhstar, degraaff_a_remarkable, DeGraaff_BHstar, Begelman_bh_star, barro_from, sun_bhstar_host}, line broadening does not arise from gas kinematics, as in the classical AGN paradigm involving the BLR, but rather from interactions between photons and dense layers of gas surrounding the central BH. The analysis of \citet{Chang_lines_bhstar} accounted for several physical processes, including resonance scattering of Balmer lines, Raman scattering, and Thomson scattering. However, several of our findings appear inconsistent with the possibility that the BH$^*$ model (as currently proposed) would be enough to explain the entire LRD population. The absence of discontinuities, near or above standard LRD selection thresholds (see Fig. \ref{fig:broad_h_bopt}), suggests that the physical mechanism responsible for broad lines in non-LRD galaxies ($\delta_{\text{v-shape}} < 0.6$) remains present in the LRD regime ($\delta_{\text{v-shape}} > 0.6$). This reveals a clear evolutionary or physical link between non-V-shaped broad H$\alpha$ emitters, typically classified as regular AGNs, and LRDs. Since the broad lines in the AGN paradigm arise from virialized motion within the BLR, the BH mass is directly linked to the width of the observed lines through the relation, $M_\text{BH} \propto R \Delta V^2 /G$, \citep[e.g.,][]{Greene_BH_mass, Reine_BH_AGN}. While the BH$^*$ interpretation invokes purely interactions with dense gas, a fundamental dissonance emerges between these two models for what appears to be a single, continuous population. Again, this applies to the bulk of the LRD population and does not preclude the possibility that some LRDs present unique features or the presence of a pure BH$^*$.

As discussed in Sect. \ref{sec:BBK}, non-stellar Balmer breaks are exclusively observed in compact, V-shaped galaxies but does not account for the majority. In \citet{DeGraaff_BHstar}, the strength of these non-stellar Balmer breaks was modeled using a modified blackbody with $T_{\text{BB}} \approx 5000$ K. Combined with the absorption features detected in several LRDs \citep[e.g.,][]{kocevski_rise_2024, matthee_little_2024, barro_from}, it suggests the presence of a high-density gas reservoir along the line of sight \citep{Inayoshi_2025, deugenio_spectra, degraaff_a_remarkable, Naidu_bhstar} in several LRDs. Given the continuous transition toward the LRD state identified in our work, a corresponding continuity in terms of the physical processes is expected among AGNs and LRDs.

\begin{figure}[h!]
    \centering
    \includegraphics[width=9cm]{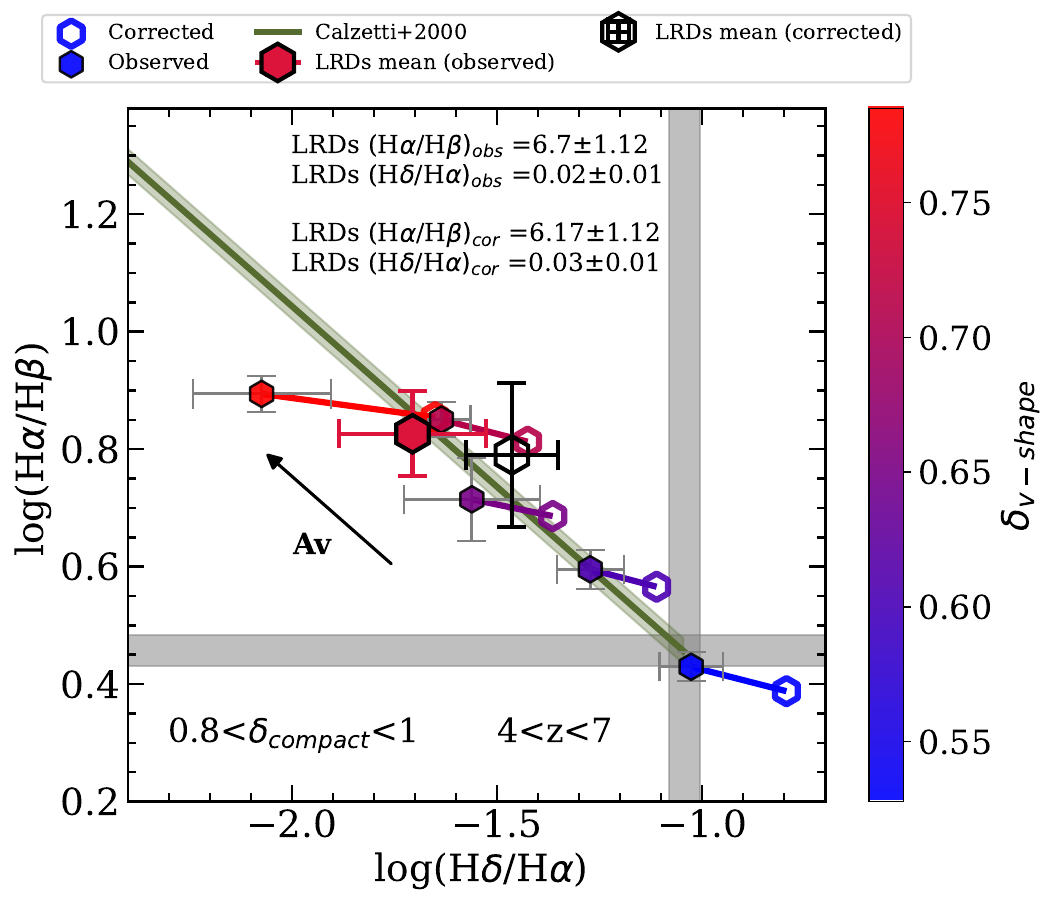}
    \caption{H$\alpha$/H$\beta$ as a function of H$\delta$/H$\alpha$, color coded by the intensity of the v-shape, $\delta_\text{v-shape}$. The filled hexagons represent the observed ratios whereas open ones illustrate the impact of corrected Balmer absorption lines. Both observed and corrected ratios are linked by a continuous line for clarity. Case B recombination is highlighted for both ratios as a gray thick line. The predicted dust ratios using the Calzetti extinction law are shown as a dark green line, with error bars corresponding to the intrinsic Case B scatter ($5\%$).}
    \label{fig:balmer_lines_ratio}
\end{figure}

This continuity, combined with the agreement between the observed Balmer ratios and a dust-attenuated case~B recombination (see Fig.~\ref{fig:balmer_lines_ratio}), suggests that the majority of LRDs likely harbor a attenuated BLR responsible for the observed line broadening, analogous to those found in standard Type~1 AGNs. This interpretation further implies that dense gas is unlikely to be the sole physical process responsible for the line broadening or the reddening observed in LRDs. While the dense gas reservoir invoked in the BH$^*$ framework may apply to specific LRD sources, these configurations are likely transient. Over time, such systems might either self-disrupt or evolve kinematically through interactions with the central BH and its environment. This process would render extreme configurations short-lived, consistent with our findings that more extreme sources exhibit shorter lifetimes and duty cycles (see Sect.~\ref{subsec:corralation_dvshape_dcompact}). This evolution could ultimately lead to the fragmentation of the gas into dense clouds, forming the BLR, while simultaneously reducing the overall gas density in the surrounding environment. This reduction in density would, in turn, result in Balmer break strengths consistent with those observed in typical AGNs \citep{Inayoshi_2025}.
A recent model proposed by \citet{madau_maiolino_LRDs} reinforces this perspective, interpreting the LRD population as Type 1 AGNs observed within a specific range of geometric configurations and inclination angles. Under this framework, LRDs do not constitute a distinct class of objects but rather represent the extreme tail of the Type 1 AGN distribution, featuring a BLR composed of dense gas clumps.
This is further supported by recent discoveries of LRDs transitioning into typical AGNs, as well as X-ray detections in several sources \citep{kocevski_rise_2024,Fu_LRD_xray, Hviding_XRD}. These detections suggest that the gas has become sufficiently diffuse along the line of sight to allow for X-ray photons to escape, thereby probing a lower density environment than previously observed.

\section{Conclusion}

In this work, we investigate whether LRDs represent a distinct class of astronomical objects or whether they are instead congruous with respect to the broader galaxy population. To this end, we combined photometric, spectroscopic, and morphological data over $\approx 750$ arcmin$^2$, then introduced a continuous metric to evaluate the defining features of LRDs. This approach allowed us to avoid arbitrary selection effects, which might otherwise introduce a biased view of the population. \\

A clear connection emerges between their three primary features: morphology, V-shaped SED prominence, and broad-line strength. Specifically, as the V-shaped SED becomes more pronounced, the morphology becomes increasingly compact, which correlates with a rising contribution from the broad emission line component. Crucially, we find no discontinuities between the conventionally selected LRD population and the general galaxy population. Instead, we observe a smooth transition between the intensity of the V-shape and both the fraction of compact galaxies (see Fig. \ref{fig:proportion_compact_v_shape}-left) and the broad-line contribution to the H$\alpha$ complex (see Fig. \ref{fig:broad_h_bopt}). This suggests that LRDs are not a separate class of objects, but rather the extreme end of a continuum defined by increasing compactness, V-shape intensity, and line broadening. \\

Furthermore, extreme Balmer breaks (exceeding 3) are predominantly found in the most extreme LRD states. However, these account for only $\approx 3\%$ of the sample, suggesting that such rare sources represent a short-lived evolutionary phase and contribute to the overall heterogeneity of the population. \\

Little red dots present a deficit in [N\,\textsc{ii}] but we find that this deficit is not restricted to these sources (see Fig. \ref{fig:NII}). Instead, it appears to be a global property of compact, metal-poor sources, indicating a low mass host. The observed continuity in the properties of broad H$\alpha$ lines, combined with the consistency of the Balmer line ratios with dust-attenuated Case B recombination, suggests that the physical mechanisms responsible for line broadening in LRDs are consistent with the presence of a dusty BLR, in line with the standard AGN paradigm. However, we note we cannot exclude secondary contributions, such as dense gas in the line of sight. \\

As another sign of similarity, the continuity between LRD and non-LRD galaxies in the intensity of their Balmer decrement suggests that the red color of LRDs is dominantly due to dust attenuation. Since current observations cannot definitively exclude the presence of dust, its potential role should not be dismissed. Assuming the Balmer decrement is solely produced by stellar attenuation, we can infer an upper limit for the dust mass in LRDs of $M_\text{dust max, LRDs} \approx 3 \times 10^{4}~M_\odot$.\\

\begin{acknowledgements}
      This work is based on observations made with the NASA/ESA/CSA \textit{James Webb} Space Telescope (JWST). The data were obtained from the Mikulski Archive for Space Telescopes at the Space Telescope Science Institute, which is operated by the Association of Universities for Research in Astronomy, Inc., under NASA contract NAS 5-03127 for JWST. These observations are associated with JWST programs ERS-1345 (CEERS), GO-1837 (PRIMER), GO-2561 (UNCOVER), GO-4233 (RUBIES), as well as GTO programs 1180, 1181, 1210, 1215, 1286, 1287 (JADES). We also acknowledge the DAWN JWST Archive (DJA), for providing the reduced data products and photometric catalogs used in this analysis.
      This project has received funding from the European Union’s Horizon 2020 research and innovation programme under the Marie Skłodowska-Curie grant agreement No 101148925. We would acknowledge the following open source software used in the analysis: \texttt{Astropy} \citep{Astropy} , \texttt{photutils} \citep{photutils}, \texttt{Numpy} \citep{numpy}.
   
\end{acknowledgements}

\bibliographystyle{aa}
\bibliography{LRDs}

\begin{appendix}

\section{On the H$\alpha$ contribution to the observed broadband flux}\label{sec:appendix0}

One of the potential biases in photometric studies, and consequently in the measurement of the  $\delta_\text{v-shape}$ parameter, arises from the boosting effect of emission lines within broad filters. This is particularly relevant for LRDs when computing the optical slope, $\beta_\text{opt}$, as discussed in \citet{kocevski_rise_2024}. The H$\alpha$ line flux can boost the last filter used to compute the optical slope, and therefore we can overestimate the redness of LRDs in photometric studies. To quantify this bias, we utilize the public NIRSpec MSA v4.4 table available on the DJA (see Sect. \ref{spectroscopic_data}). We focus on PRISM spectra where the H$\alpha$ line falls within the red broad bands of the JWST, specifically in the F277W, F356W, or F444W filters. We then measure the H$\alpha$ flux for approximately 5,000 sources in the corresponding filter by employing
\begin{eqnarray}
      F_{\text{H}\alpha,\text{filter}} = \frac{\int f_{\lambda, \text{H}\alpha} \ T(\lambda) \ \text{d}\lambda}{\int  \ T(\lambda) \ \text{d}\lambda} ,
\end{eqnarray}

\noindent where $T(\lambda)$ represents the transmission curve of the filter, and $F_{\text{H}\alpha,\text{filter}}$ denotes the flux of the H$\alpha$ line measured by the corresponding filter. To isolate the H$\alpha$ line ($f_{\lambda,\text{H}\alpha}$) from the PRISM data, we first subtract the continuum using a power-law fit. We then extract the spectral region within a $0.2~\mu$m window centered on the H$\alpha$ emission. This isolated line profile is resampled onto a broader wavelength grid that matches the transmission profile of the corresponding filter, after which we apply Eq.~(2).

\begin{figure}[h!]
    \centering
    \includegraphics[width=6.5cm]{ 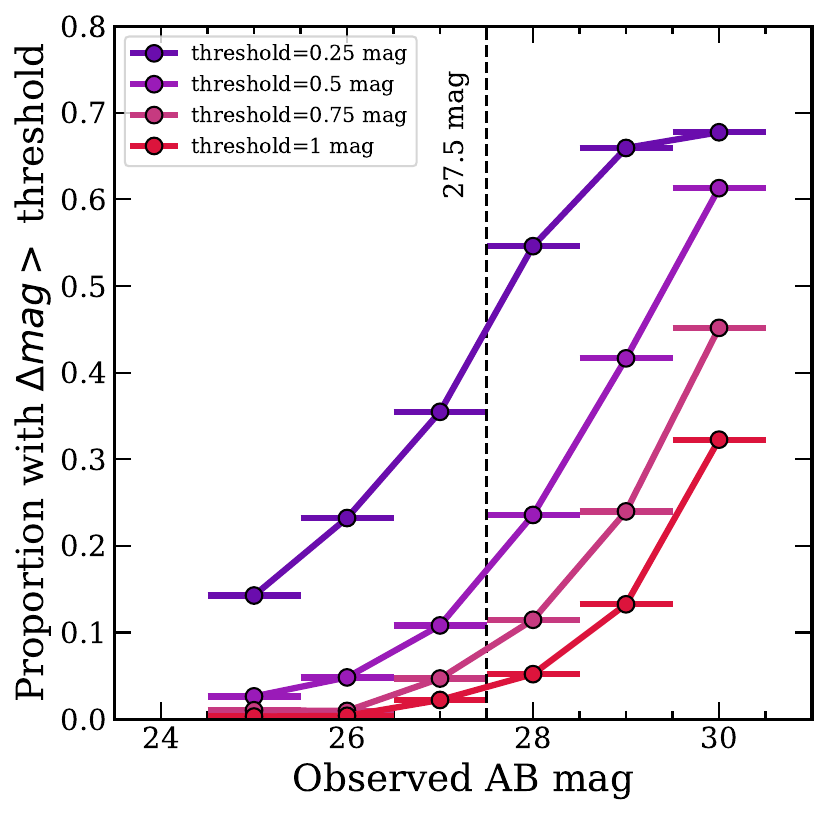}
    \caption{Illustration of the H$\alpha$ boosting effect: the proportion of galaxies for which the H$\alpha$ line in PRISM spectra coincides with either the F277W, F356W, or F444W filters, requiring a correction in AB magnitude above specific thresholds (from 0.25\,AB mag in purple to 1\,AB mag in red), is plotted as a function of the observed AB magnitude.The vertical dashed line denotes a magnitude of 27.5 AB. Below this limit, fewer than 20$\%$ of the galaxies require a correction exceeding 0.5 mag. 
}
    \label{fig:contribution_percent_halpha_mag}
\end{figure}

Figure~\ref{fig:contribution_percent_halpha_mag} illustrates the fraction of sources requiring a correction in AB magnitude above fixed values (0.25, 0.5, 0.75, and 1 mag) as a function of the observed AB magnitude (in F277W, F356W, or F444W). The contribution of H$\alpha$ to the observed flux is inversely related to the source brightness: at 28 AB mag, approximately 55\% of galaxies observed in filters covering H$\alpha$ require a correction exceeding 0.25 AB mag. This fraction decreases gradually for brighter sources. As expected, fewer galaxies require larger corrections, only 25\% at 28 AB mag need a correction above 0.5 mag, and less than 10\% require a correction above 1 mag.

We conclude that applying a threshold of 27.5 AB mag in broad-band filters covering H$\alpha$ (here after denoted $m_{\text{H}\alpha}$) ensures minimal contamination from line emission: fewer than 20\% of galaxies will require a correction above 0.5 mag, and fewer than 10\% will need a correction above 0.75 mag. We caution that this analysis does not account for the magnitude limits of individual fields, and variations in this threshold are expected for shallower or deeper surveys. Moreover, extended galaxies might need a more important correction due to slit-loss. Nevertheless, this provides a quantitative estimate of the H$\alpha$ boosting effect. Consequently, LRDs fainter than 27.5 AB mag should not be selected using photometric methods, as their red colors may be significantly contaminated by this effect. All along this work we therefore apply the following threshold: we limit the magnitude of the filter containing H$\alpha$ to be brighter than 27.5 AB mag, $m_{\text{H}\alpha}$<27.5 AB mag. Applied to a subsample of 41 LRDs (from \citealt{kocevski_rise_2024} sample), the correction applied to the optical slope $\beta_\text{opt}$ is on average 0.18.

\section{Compactness with aperture photometry}\label{sec:appendix1}

\begin{figure}[h]
    \centering
    \includegraphics[width=8cm]{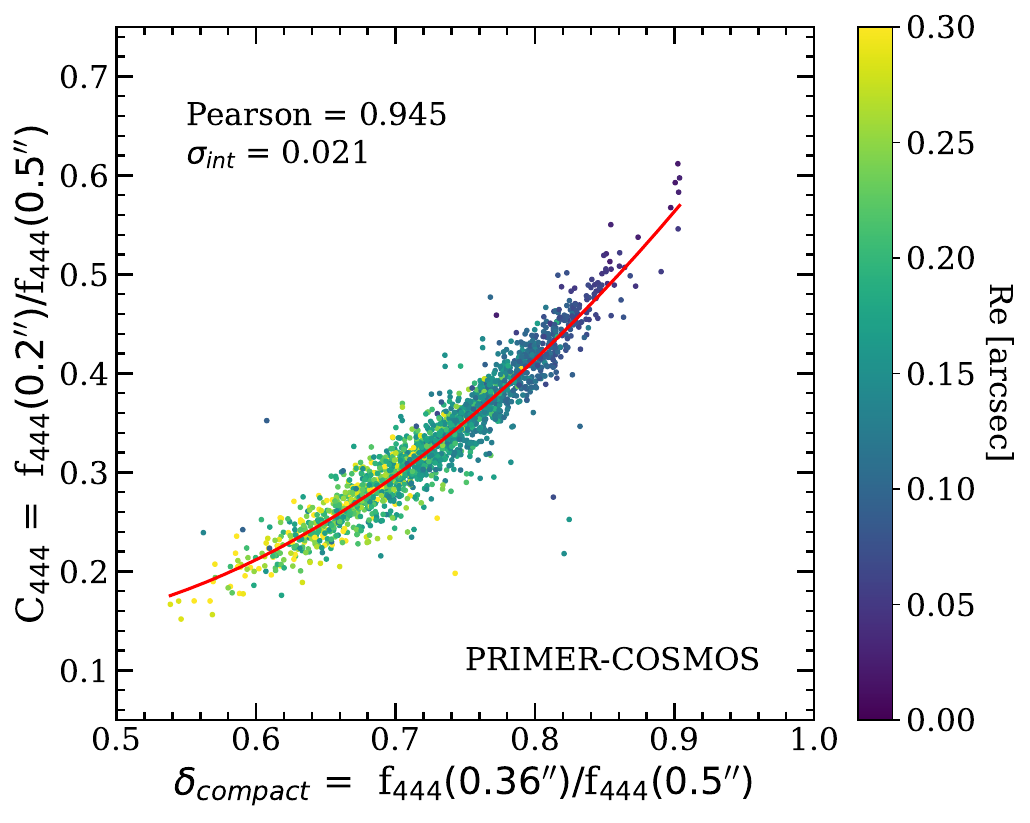}
    \caption{ $C_{444}$ as a function of $\delta_\text{compact}$. The red curve corresponds to the best polynomial fit, while the color map indicates the effective radius ($R_e$) from the 2D S\'ersic fit provided by the DJA \citep{Genin_DJA}. Both the Pearson correlation coefficient and the intrinsic scatter of the best fit are also shown.}
    \label{fig:aperture_comparaison}
\end{figure}

In this section we compare the aperture photometry ratios to ensure that the $\delta_\text{compact}$ parameter, defined as the ratio $f_{444}(0.36'') / f_{444}(0.5'')$, is capable of effectively probing compact objects, similar to the $C_{444}$ parameter used in \citet{akins_cosmos-web_2024}, which is defined as $f_{444}(0.2'') / f_{444}(0.5'')$. In Fig. \ref{fig:aperture_comparaison} we focus on PRIMER-COSMOS, for which the $C_{444}$ values are provided by the COSMOS-Web catalog \citep{Shuntov_cosmos_catalog}. Both parameters are strongly correlated with small intrinsic scatter implying that, like $C_{444}$, $\delta_\text{compact}$ is an efficient proxy for identifying compact sources.

\section{Sigmoid function}\label{sec:appendix2}

\begin{figure}[h]
    \centering
    \includegraphics[width=7cm]{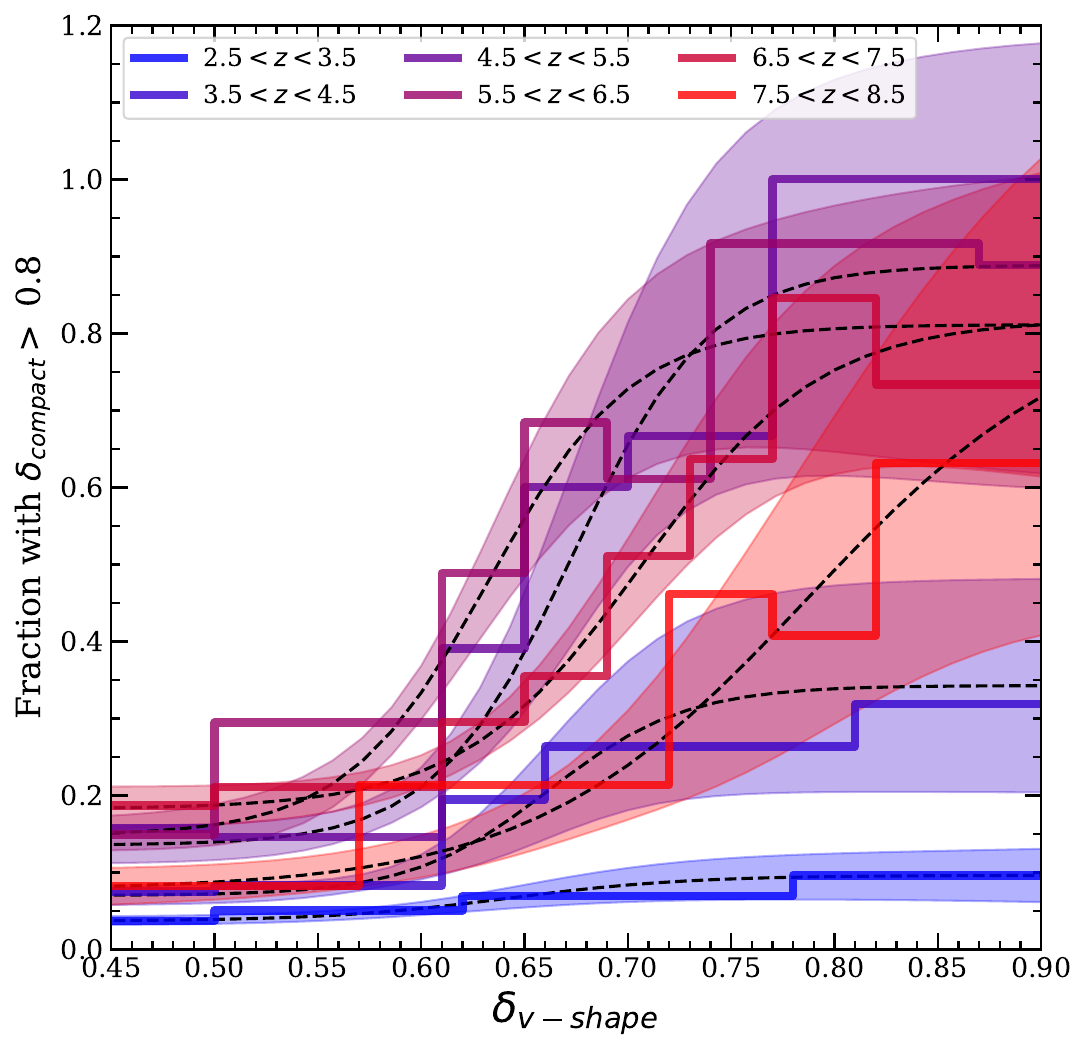}
    \caption{Fraction of galaxies with $\delta_\text{compact} > 0.8$ as a function of $\delta_\text{v-shape}$, shown for redshift bins ranging from $z = 3 \pm 0.5$ to $z = 8 \pm 0.5$. The solid lines represent the observed data, while the dashed lines indicate the best-fit sigmoid models, with the associated uncertainties depicted as shaded regions. }
    \label{fig:sigmoid}
\end{figure}

To investigate the fraction of compact galaxies as a function of the V-shape intensity, $\delta_\text{v-shape}$, and redshift, we observe that the relationship at each redshift can be effectively described by a sigmoid function. We therefore fit the data using the following formula:

\begin{equation}
y = \frac{L}{1 + e^{-k(x - x_0)}} + b,
\end{equation}

\noindent where $x$ represents $\delta_\text{v-shape}$, $x_0$ is the median value of the sample ($\delta_\text{v-shape, median}$), and $y$ corresponds to $\delta_\text{compact}$. In Fig.~\ref{fig:sigmoid} the dashed lines represent the best-fit sigmoid curves for each redshift bin, while the shaded regions indicate the uncertainties in the fit, derived through a Monte Carlo approach.

\section{BIC comparison}\label{sec:appendix3}
\begin{figure}[h]
    \centering
    \includegraphics[width=7cm]{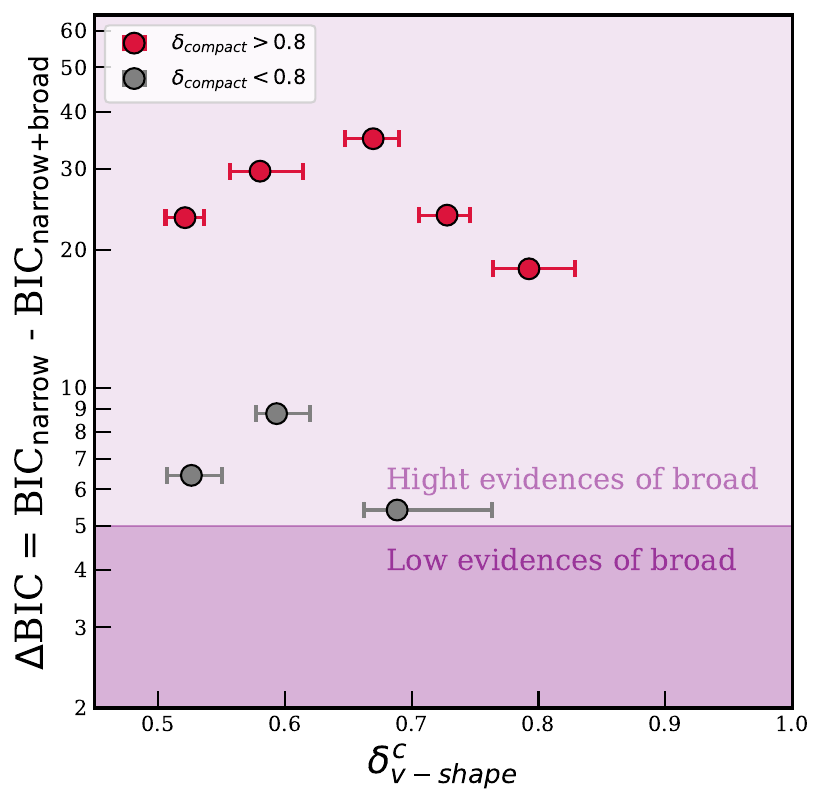}
    \caption{Comparison of BIC for models with and without a broad component. Compact sources are represented by red circles, while extended sources are shown as gray circles.
}
    \label{fig:BIC_comparison}
\end{figure}

To assess the presence of broad lines, we utilized the BIC, defined as

\begin{equation}
\text{BIC} = \chi^2 + k \times \log(n),
\end{equation}

\noindent where $k$ is the number of free parameters in the model, and $n$ is the number of data points. We computed the BIC for models both with and without a broad component. Figure~\ref{fig:BIC_comparison} illustrates the difference between the two BIC values, $\Delta\text{BIC}$, for our compact sample in red ($\delta_\text{compact} > 0.8$) and our extended sample in gray ($\delta_\text{compact} < 0.8$). As detailed in Sect.~\ref{sec:Halpha_evolution}, we consider a broad line to be reliably detected and necessary when $\Delta\text{BIC} > 5$ and the S/N of the broad component exceeds 5.

\clearpage
\section{H$\alpha$ line fit}\label{sec:appendix4}

\vspace{0.5cm} 

\begin{center}
    \begin{minipage}{\textwidth} 
        \centering
        \includegraphics[width=16cm]{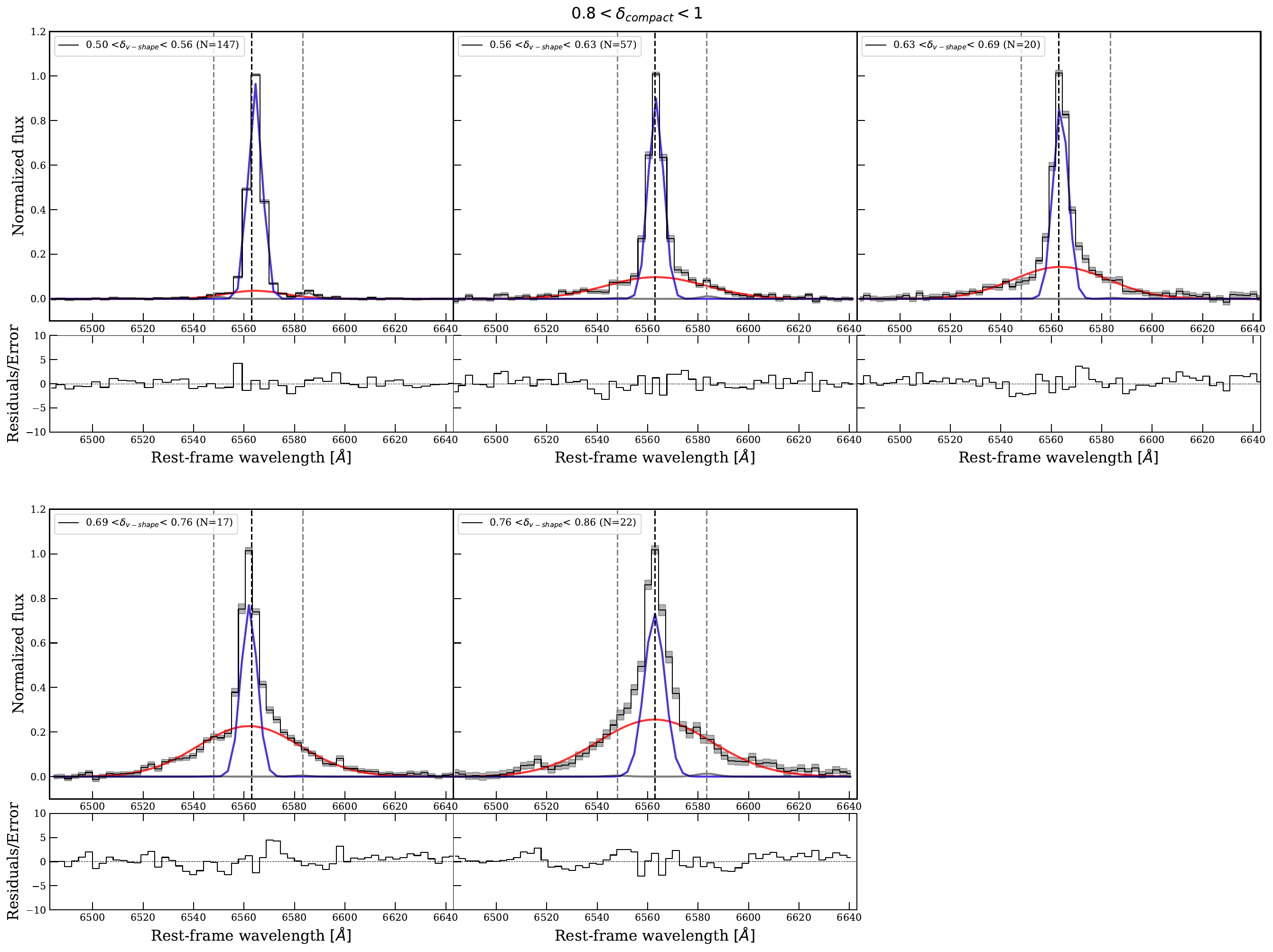}
        \captionof{figure}{Results of the H$\alpha$ complex fit for the compact galaxy group ($0.8<\delta_{\text{compact}} < 1$). The broad H$\alpha$ component is shown in red, while the narrow component is in blue. The [N\,\textsc{ii}] doublet is also displayed in solid light gray, for each bin of $\delta_{\text{v-shape}}^c$.}
        \label{fig:all_lines_compact}
    \end{minipage}
\end{center}

\vspace{2cm}

\begin{center}
    \begin{minipage}{\textwidth} 
        \centering
        \includegraphics[width=16cm]{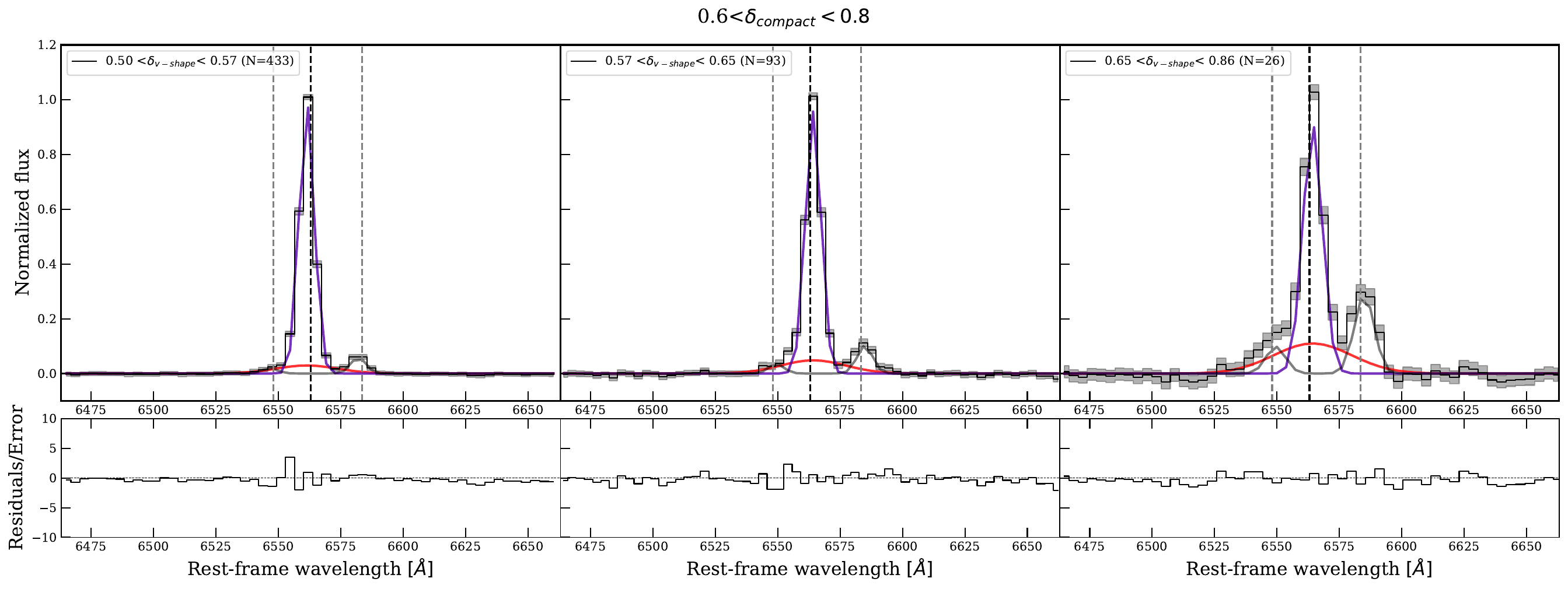}
        \captionof{figure}{Same as Fig. \ref{fig:all_lines_compact} but for extended sources ($0.6<\delta_\text{compact} < 0.8$).}
        \label{fig:all_lines_extended}
    \end{minipage}
\end{center}

\end{appendix}

\end{document}